\documentclass{rsproca_new}%%%%where rsproca is the template name

\usepackage{fixltx2e}
\usepackage{physics}
\usepackage{wrapfig}
\usepackage{makecell}
\usepackage{gensymb}
\usepackage[caption=false]{subfig}

\newcommand{\si}[1]{#1}
\newcommand{\angstrom}{\mbox{\normalfont\AA}}

\begin{document}

%%%% Article title to be placed here
\title{Computation of Burgers Vectors from Elastic Strain and Lattice Rotation Data}

\author{%%%% Author details
J. Cloete$^{1}$, E. Tarleton$^{1,2} $and F. Hofmann$^{1}$ } 

%%%%%%%%% Insert author address here
\address{$^{1}$Department of Engineering Science, University of Oxford, Parks Road, Oxford OX1 3PJ, UK\\
$^{2}$Department of Materials, University of Oxford, Parks Road, Oxford OX1 3PH, UK}

%%%% Subject entries to be placed here %%%%
\subject{Mechanics, Materials Science,Computer Modelling and Simulation, Dislocations}

%%%% Keyword entries to be placed here %%%%

%%%% Insert corresponding author and its email address}
\corres{J. Cloete, E. Tarleton, F. Hoffmann \\
\email{jacques.cloete@eng.ox.ac.uk \\
edmund.tarleton@eng.ox.ac.uk \\
felix.hofmann@eng.ox.ac.uk}}

%%%% Abstract text to be placed here %%%%%%%%%%%%
\begin{abstract}

A theoretical framework for computation of Burgers vectors from strain and lattice rotation data in materials with low dislocation density is presented, as well as implementation into a computer program to automate the process. The efficacy of the method is verified using simulated data of dislocations with known results. A 3D data set retrieved from Bragg coherent diffraction imaging (BCDI) and a 2D data set from high resolution transmission Kikuchi diffraction (HR-TKD) are used as inputs to demonstrate the reliable identification of dislocation positions and accurate determination of Burgers vectors from experimental data. For BCDI data, the results found using our approach show very close agreement to those expected from empirical methods. For the HR-TKD data the predicted dislocation position and the computed Burgers vector showed fair agreement with the expected result, which is promising considering the substantial experimental uncertainties in this dataset. The method reported in this paper provides a general and robust framework for determining dislocation position and associated Burgers vector, and can be readily applied to data from different experimental techniques.
\end{abstract}
%%%%%%%%%%%%%%%%%%%%%%%%%%%

%%%%%%%%%% Insert the texts which can accomdate on firstpage in the tag "fmtext" %%%%%

\begin{fmtext}

\end{fmtext}

%%%%%%%%%%%%%%% End of first page %%%%%%%%%%%%%%%%%%%%%

\maketitle

\section{Introduction}
%%%% Insert A head here
Crystal lattice defects such as dislocations play a key role in controlling the properties of high performance materials central to modern life. Examples include alloys for aerospace, nuclear and automotive applications \cite{gupta,guo,zhang}. Dislocations are also important in semiconductors, for example in photovoltaic devices, for they act as impurity segregation sites that are detrimental to solar cell performance \cite{hirsch1,hirsch2,ourmazd,pauls}.

To optimise material performance, a detailed understanding of dislocations is key. This is particularly important for tuning materials in order to modify dislocation behaviour \cite{anderson,bacon,cai}, e.g. by modifying the microstructural landscape of the material. Indeed this is one of the main ways modern alloys are optimised \cite{smallman,clemens,pollock,kuziak}.

Dislocations interact via the strain fields they cause \cite{anderson,bacon}. As such, measurement techniques that allow us to image strain fields are of great importance. It is possible to measure 2D strain fields surrounding dislocations at the nano-scale using high resolution transmission electron microscopy (HR-TEM) and scanning transmission electron microscopy \cite{hytch,zhao1,zhao2}, as well as high resolution transmission Kikuchi diffraction (HR-TKD) in the scanning electron microscope \cite{keller,yu}. To measure 3D strain fields and morphology of microcrystals, Bragg coherent diffraction imaging (BCDI) has achieved successful results \cite{robinson,pfeifer,hofmann}.

A key challenge when characterising dislocations is the accurate determination of the Burgers vector \cite{burgers} for specific dislocations. A classic approach is to use $\bm{g}\cdot\bm{b}$ contrast, wherein dislocations cause little diffraction contrast when the diffraction vector $\bm{g}$ is perpendicular to the Burgers vector $\bm{b}$, i.e. when the invisibility criterion $\bm{g}\cdot\bm{b} = 0$ is met \cite{bacon}. For materials with well-known crystallography, multiple $\bm{g}$-vectors are required in TEM as there are several possible directions that the Burgers vector can take, and only by finding two $\bm{g}$-vectors for which the invisibility criterion is met can the Burgers vector direction be determined. Even then, there is ambiguity to the sign of the Burgers vector as it could be positive or negative, and this ambiguity can only be resolved by detailed comparison with diffraction image simulations \cite{hofmann,yu}. There is also no immediate indication of Burgers vector magnitude from these techniques, and one must refer to the known crystallographic structure of the material to determine this. A further problem arises where the preferred Burgers vector directions are not known a priori. In this case, $\bm{g}\cdot\bm{b}$ contrast cannot be easily applied as there is no limited set of Burgers vectors to try.

Since measurements of the full lattice strain tensor are now readily possible in 2D and 3D, an attractive idea is to use the measured strain fields at the continuum scale to directly determine the Burgers vector of specific dislocations. Two methods present themselves; the first being to integrate the displacement gradient of the crystal lattice about a Burgers circuit enclosing the dislocation, the second to integrate the computed Nye tensor (proposed by Nye \cite{nye}) of the crystal across a surface through which the dislocation passes \cite{rima}.

Here, we first lay out the underlying theory and concepts used before presenting a framework for determining dislocation Burgers vector direction and magnitude from experimentally-measured strain and lattice rotation fields. We also cover the key concepts for implementation of this framework into a program. The Nye tensor approach is initially investigated, however this method turns out not to be suitable for the task at hand. Thus, the framework presented in this paper relies on integration of the displacement gradient around a Burgers circuit. Initially we consider a mathematical model of the displacement gradient surrounding an infinite straight mixed dislocation of configurable sense $\bm{\xi}$ and Burgers vector $\bm{b}$. This is used to verify the efficacy of the method and reliability of the approach as well as its sensitivity to factors such as noise. Experimental data sets from BCDI and HR-TKD techniques are then used to locate and characterise dislocations in real material specimens. Comparisons to known results found from techniques such as $\bm{g}\cdot\bm{b}$ contrast are then made. It should be noted that the focus of this paper is to compute the Burgers vector for individual dislocations in materials where dislocation densities are sufficiently low that voxel spacing is much smaller than the spacing between dislocations.
%%%%%%%%%%%%%%%%%%%%%%%%%%%%%%%%%%%%%%%%%%%%%%%%%%%%%%%%%%%%%%%%%%%%%%%%%%%%%%%%%%%%%%%
\\
\section{Theory and Methodology}
\subsection{Mathematical Definition of a Burgers Vector}

\begin{figure}
\centering
\subfloat[]{\includegraphics[width=0.3\textwidth]{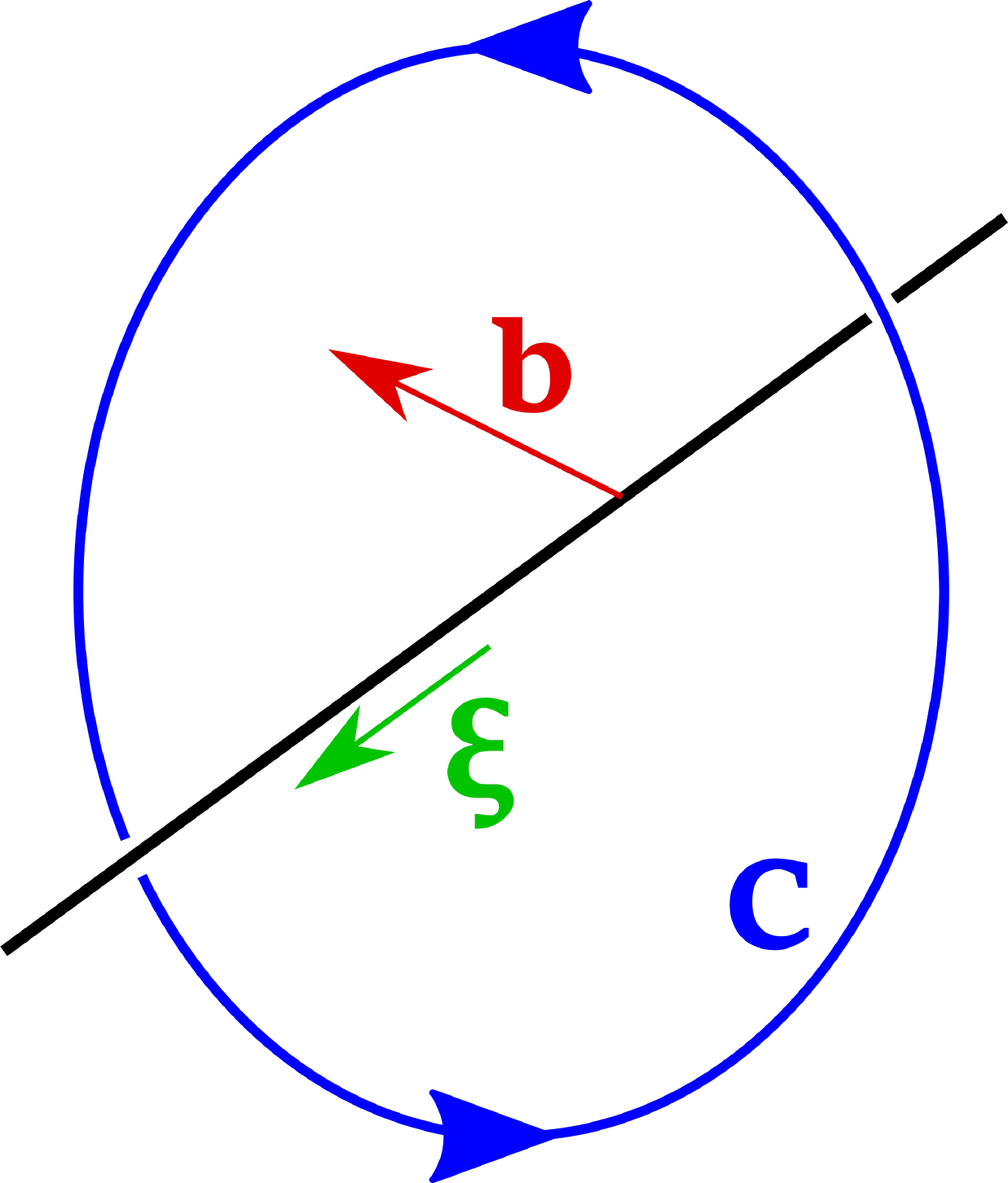}} \hspace{0.05\textwidth}
\subfloat[]{\includegraphics[width=0.6\textwidth]{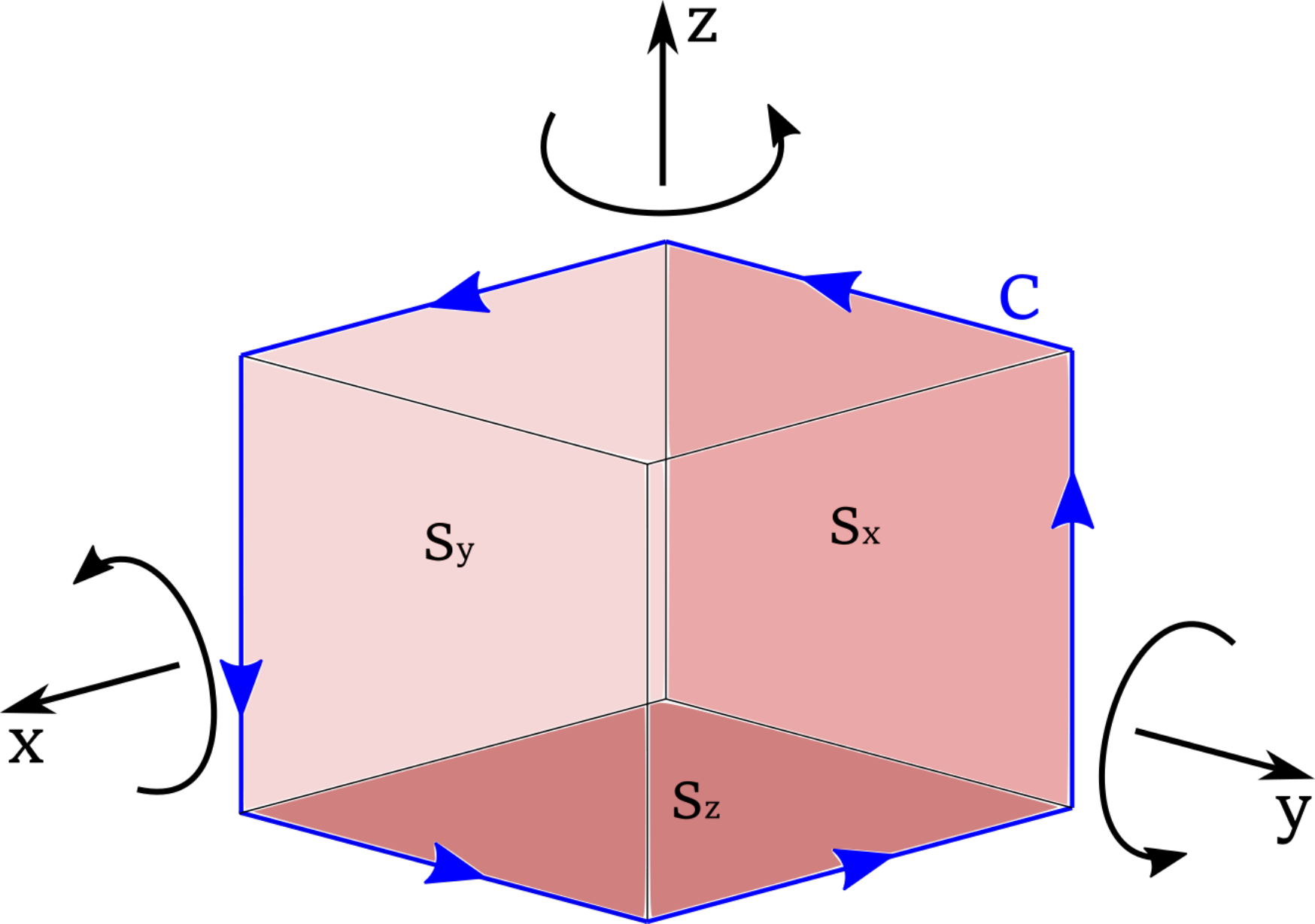}}
\caption{a) A dislocation line of Burgers vector $\bm{b}$ and direction defined by sense $\bm{\xi}$ enclosed by Burgers circuit $\bm{C}$. b) Integration loop $\bm{C}$, now in a form suitable for numerical computation of Burgers vector $\bm{b}$.}
\label{fig:IntegrationLoop3D}
\end{figure}

A Burgers vector describes the discontinuity in a material structure due to the presence of a dislocation \cite{anderson,bacon}. A mathematical definition for the Burgers vector $\bm{b}$ of a dislocation is the line integral of the elastic displacement gradient $\bm{\beta}=\bm{\nabla u}$ (where $\bm{u}$ is the elastic displacement field of the material) along the closed loop $\bm{C}$ that encloses the dislocation (also known as the Burgers circuit, defined in the original undistorted coordinate system) \cite{anderson};
\begin{align} \label{eq:BurgersVector}
\bm{b} = \oint_{C} \bm{\beta} \cdot d \bm{l}
\end{align}
where
\begin{align}
d \bm{l} = \begin{bmatrix}
dx \\
dy \\
dz
\end{bmatrix}
\end{align}
(Note that $\bm{\beta} = \bm{\varepsilon} + \bm{\omega}$, the elastic strain and lattice rotation tensor fields respectively) \cite{hlt,rima}.

This means that, as long as we know the strain and lattice rotation fields within a material, the Burgers vector $\bm{b}$ for dislocations enclosed within some Burgers circuit $\bm{C}$ can be calculated.

\subsection{The Nye Tensor Approach}
On the continuum scale, Nye defined the dislocation tensor (or `Nye tensor') $\bm{\alpha}^{Nye}$ as a geometrical relation between lattice curvature and the distribution of geometrically-necessary dislocations (GNDs); dislocations required to accommodate the plastic strain gradients in a material \cite{anderson,nye,rima}.

Nye related $\bm{\alpha}^{Nye}$ to the dislocation distribution inside a material \cite{anderson,nye,rima}; for each possible slip system $\lambda$, given $q$ dislocations per unit area each with Burgers vector $\bm{b}$ and sense $\bm{\xi}$ threading the plane,
\begin{align}
\alpha^{Nye}_{km} = q b_{k} \xi_{m}
\end{align}
Let GND density $\bm{\rho}$ be defined such that $\rho_{m} = q \xi_{m}$; therefore
\begin{align}
\bm{\alpha}^{Nye} = \sum_{i = 1}^{\lambda} (\bm{b}^{i} \otimes \bm{\rho}^{i})
\end{align}

One might argue that if $\bm{\alpha}^{Nye}$ can be found, the Burgers vector $\bm{b}$ can be evaluated using the surface integral as opposed to the line integral. An immediate benefit would be improved numerical integration, as one could integrate across an entire surface of voxels rather than just a line. The significantly increased number of data points used would hopefully reduce error due to noise and discretisation of the data.

However, upon further investigation it has been deduced that the Nye tensor approach to calculating the Burgers vector is not feasible for the applications covered in this paper. This is because, for the case of individual dislocations, $\bm{\alpha}^{Nye}$ vanishes everywhere except at the dislocation core, where it is instead undefined, meaning that the surface integral cannot be evaluated to find the Burgers vector. This is a result of the discontinuity associated with the individual dislocation being confined to the dislocation core, such that dislocation density is infinite at the core and zero elsewhere. In addition, the Nye tensor at a point essentially corresponds to having an infinitesimally small Burgers circuit at that point. As a result, the Nye tensor evaluated for the elastic distortions associated with a single dislocation is undefined at the origin and zero everywhere else. A full demonstration and explanation are provided in Appendix \ref{nye_further}.

The rest of this paper will explore the method of integration of the displacement gradient around a Burgers circuit to compute the Burgers vector.

\subsection{Numerical Computation}
Elastic strains and lattice rotations obtained experimentally are measured as discrete voxels of data rather than being defined by algebraic functions (as with the mathematical model). Therefore, rather than integrating analytically, we must make use of numerical techniques.

 To compute Burgers vector $\bm{b}$ we perform the line integral in Eq.(\ref{eq:BurgersVector}). Note that, in Cartesian coordinates, $\bm{\beta} \cdot d \bm{l} $ can be written as
\begin{align}
\bm{\beta} \cdot d \bm{l} = 
\begin{bmatrix}
\beta_{11} & \beta_{12} & \beta_{13} \\
\beta_{21} & \beta_{22} & \beta_{23} \\
\beta_{31} & \beta_{32} & \beta_{33}
\end{bmatrix}
\cdot
\begin{bmatrix}
dx \\
dy \\
dz
\end{bmatrix}
=
\begin{bmatrix}
\beta_{11}dx + \beta_{12}dy + \beta_{13}dz \\
\beta_{21}dx + \beta_{22}dy + \beta_{23}dz \\
\beta_{31}dx + \beta_{32}dy + \beta_{33}dz
\end{bmatrix}
\end{align}
Therefore, integrating $\bm{\beta}$ with respect to $\bm{l}$ requires integrating each of the nine components of $\bm{\beta}$ with respect to either $x$, $y$ or $z$. The limits for, and equations relating, these three variables are determined by the choice of Burgers circuit $\bm{C}$.

The simplest Burgers circuit design that will be effective for dislocations in any orientation in 3D space is the loop $\bm{C}$ shown in Fig. \ref{fig:IntegrationLoop3D} b). The Burgers circuit encloses a cube (or cuboid), and is designed such that a right-hand screw dislocation pointing in the direction of a positive coordinate axis gives a positive result for $\bm{b}$ (this should be borne in mind when considering the signs of computed Burgers vector directions).

For integration along loop $\bm{C}$ shown, along each of the six straight lines, two of $dx$, $dy$ or $dz$ will be zero. Thus the line integral of displacement gradient $\bm{\beta}$ is reduced to a total of 18 integrals of one variable (six for each of the $x$-, $y$- and $z$-components of $\bm{b}$) that can each be computed individually. This is ideal for numerical integration.

For data that is discrete and exists across a set of voxels, we can define a Burgers circuit as described above that passes through the centres of a chain of voxels to form integration loop $\bm{C}$. The discrete displacement gradient can then be integrated numerically, for example by using MATLAB's \textit{trapz} function or a variation thereof, and travelling along this loop of voxels. This numerical approach was used to produce a MATLAB program that computes Burgers vectors from arrays of 3D or 2D elastic strain and lattice rotation data.The specific numerical method used for the results in this paper was an implementation of Simpson's rule with error of order O($h^{4}$). Further information on implementation can be found in appendix \ref{implementation}.
%%%%%%%%%%%%%%%%%%%%%%%%%%%%%%%%%%%%%%%%%%%%%%%%%%%%%%%%%%%%%
\\
\section{Tests - Mathematical Models}
\subsection{Test Methodology}

\begin{figure} [t] \label{ExampleTestFig}
\includegraphics[width=0.7\textwidth]{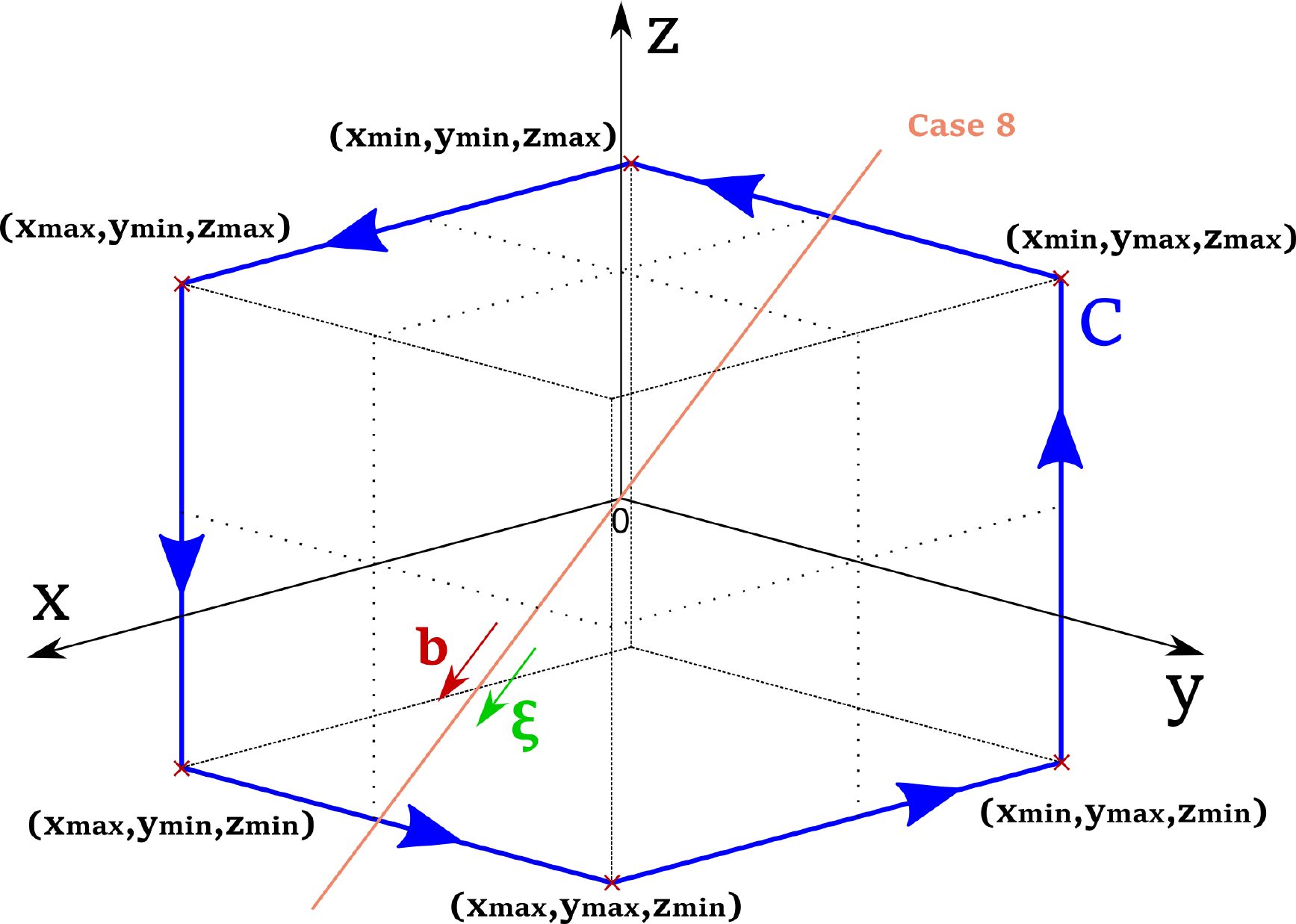}
\centering
\caption{Integration loop $\bm{C}$, now with vertex coordinates defined by the coordinate limits in the $x$, $y$ and $z$ directions. The dislocation line for Case 8 is also present, passing through the origin and with $\bm{b}$ and $\bm{\xi}$ as shown. Note that in this particular case the dislocation line passes through the edges of the integration loop, which causes an error in computation.}
\label{fig:ExampleTestFig}
\end{figure}

To test the reliability of Burgers vector determination, we initially consider a mathematical model of an infinite, straight, mixed dislocation. Nine cases were chosen and used to confirm that the program can accurately compute the Burgers vector for dislocations of various orientations. The vertices of Burgers circuit $\bm{C}$ are defined by the coordinate limits shown in Fig. \ref{fig:ExampleTestFig}.

A set of parameters must be chosen not only to generate the mathematical model of the dislocation, but also to prepare the program for the specific set of input data, as the size of the voxel spacing must be taken into account for numerical integration.

\begin{samepage}
The parameters for the test cases were as follows:
\begin{itemize}
    \item Burgers Vector Magnitude, $b = 1$ \si{\angstrom}
    \item Voxel Size $= 5$ \si{nm} (cube)
    \item Poisson's Ratio, $\upsilon = 0.3$
    \item Noise Coefficient, $\eta = 0$
    \item Burgers Circuit limits = [-102.5 \si{nm}, 102.5 \si{nm}] (expect for Case 9)
\end{itemize}
\end{samepage}

Note that the voxel spacing and coordinate limits used were such that the Burgers circuit $\bm{C}$ was defined around a cube with a side-length of 42 pixels and centred on the origin. Each of the dislocations also passed through the origin. The voxel size, Burgers vector magnitude and Poisson's ratio were chosen to be realistic to those found in experimental data sets \cite{hofmann, yu}.

\begin{samepage}
The following cases were considered, where the angles $\alpha$, $\psi$, $\theta$ and $\phi$ define the directions of Burgers vector $\bm{b}$ and sense $\bm{\xi}$, as depicted in Fig. \ref{fig:rotations} in appendix \ref{derivations}\ref{rotating}:
\begin{enumerate}
    \item[1.] Z-Screw; $\bm{\xi} = \bm{e_{z}}$, $\bm{b} = b\bm{e_{z}}$ ($\alpha = 0\degree$ , $\psi = 0\degree$, $\theta = 0\degree$, $\phi = 0\degree$)
    \item[2.] X-Screw; $\bm{\xi} = \bm{e_{x}}$, $\bm{b} = b\bm{e_{x}}$ ($\alpha = 0\degree$ , $\psi = 0\degree$, $\theta = 90\degree$, $\phi = 0\degree$)
    \item[3.] Y-Screw; $\bm{\xi} = \bm{e_{y}}$, $\bm{b} = b\bm{e_{y}}$ ($\alpha = 0\degree$ , $\psi = 0\degree$, $\theta = 90\degree$, $\phi = 90\degree$)
    \item[4.] Z-Edge; $\bm{\xi} = \bm{e_{z}}$, $\bm{b} = b\bm{e_{x}}$ ($\alpha = 90\degree$ , $\psi = 0\degree$, $\theta = 0\degree$, $\phi = 0\degree$)
    \item[5.] Mixed 1 ($\alpha = 45\degree$ , $\psi = 90\degree$, $\theta = 90\degree$, $\phi = 45\degree$)
    \item[6.] Mixed 2 ($\alpha = 45\degree$ , $\psi = 0\degree$, $\theta = 0\degree$, $\phi = 45\degree$)
    \item[7.] Sum of Three Screws (Z-Screw + X-Screw + Y-Screw)
    \item[8.] Error Case ($\alpha = 0\degree$ , $\psi = 0\degree$, $\theta = 135\degree$, $\phi = 0\degree$)
    \item[9.] Error Case ($\alpha = 0\degree$ , $\psi = 0\degree$, $\theta = 135\degree$, $\phi = 0\degree$), new z-axis limits = [-122.5 \si{nm}, 122.5 \si{nm}]
\end{enumerate}
\end{samepage}

\begin{figure}
\centering
\subfloat[]{\includegraphics[width=0.3\textwidth]{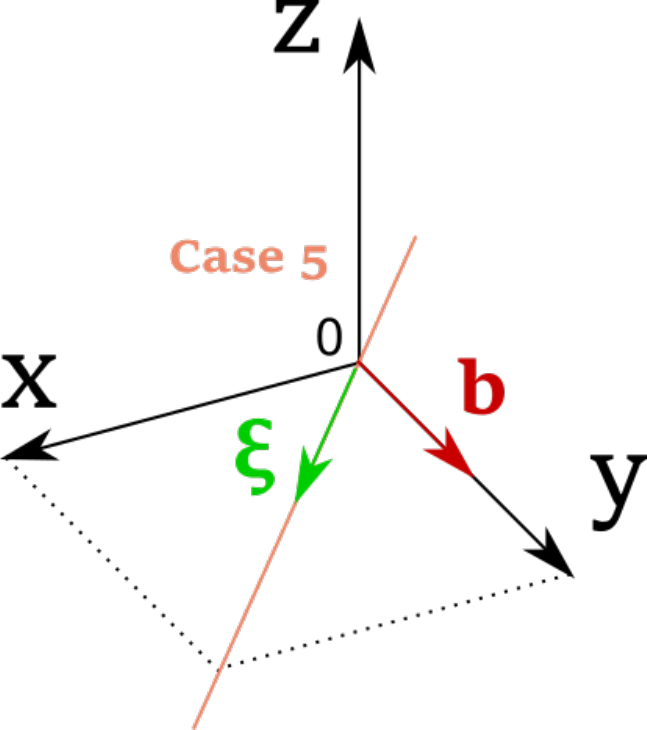}} \hspace{0.15\textwidth}
\subfloat[]{\includegraphics[width=0.3\textwidth]{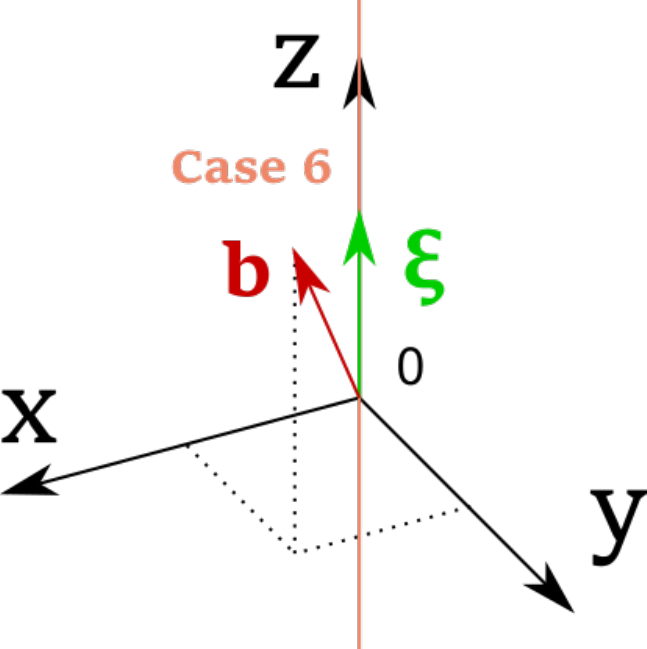}}
\caption{A visualisation of Cases 5 (a) and 6 (b), each with the dislocation line, sense $\bm{\xi}$, and Burgers vector $\bm{b}$ shown.}
\label{fig:5and6}
\end{figure}

Cases 1 through 4 as well as 7 should be straightforward to visualise, and visualisations of Cases 5 and 6 are provided in Fig. \ref{fig:5and6} and Case 8 in Fig. \ref{fig:ExampleTestFig}.

The robustness of the program was also tested. Using Case 7 as a basis, three tests were performed. In Test 1, the Burgers circuit side length was decreased, two voxels at a time, from 42 to 2. In Test 2, the fractional distance of the dislocation line from the Burgers circuit centre was increased from 0 (centre) to 1 (Burgers circuit edge). In Test 3, the noise coefficient $\eta$ was exponentially increased from $1\times10^{-6}$ to $7.86\times10^{-4}$ and the number of concentric Burgers circuits used for the removal of outliers (with $x$-, $y$- and $z$-components beyond a certain number of standard deviations from the mean values) and computation of a mean average Burgers vector was increased from 1 to 10. This average Burgers vector was computed by separately averaging the $x$-, $y$- and $z$-components from the results of the integrals around the concentric Burgers circuits. For each of the tests, the change in computed Burgers vector is reported as the angular error and percentage magnitude error compared to the ground truth.

Experimental noise/uncertainty was modelled using a standard Gaussian distribution multiplied by the noise coefficient $\eta$ and superposed onto each element of the strain and lattice rotation tensors. The distribution was truncated to within the limits of 1 standard deviation, as otherwise unrealistically large magnitudes of noise may appear in the data. The noise applied to one element of data was independent to the noise applied to adjacent elements of data.

\subsection{Results}
\begin{table}
\begin{tabular}{||m{1.7em}|m{5.4em}|m{4.1em}|m{6.6em}|m{4.7em}|m{5.1em}|m{4.7em}||}
\hline
Case No. & True Magnitude (\si{\angstrom}) & True Direction & Computed Magnitude (\si{\angstrom}) & Computed Direction & Magnitude Percentage Error (\%) & Angular Error (\degree) \\ 
\hline\hline
1. & 1 & \makecell[c]{0 \\ 0 \\ 1} & 1.0000 & \makecell[c]{0 \\ 0 \\ 1} & 2.4620e-6 & 0.0000 \\
\hline
2. & 1 & \makecell[c]{1 \\ 0 \\ 0} & 1.0000 & \makecell[c]{1 \\ 0 \\ 0} & 2.4620e-6 & 0.0000 \\
\hline
3. & 1 & \makecell[c]{0 \\ 1 \\ 0} & 1.0000 & \makecell[c]{0 \\ 1 \\ 0} & 2.4620e-6 & 0.0000 \\
\hline
4. & 1 & \makecell[c]{1 \\ 0 \\ 0} & 1.0000 & \makecell[c]{1 \\ 0 \\ 0} & 2.4620e-6 & 0.0000 \\
\hline
5. & 1 & \makecell[c]{0 \\ 1 \\ 0} & 1.0000 & \makecell[c]{0.0000 \\ 1.0000 \\ 0} & 7.4046e-6 & 3.0783e-6 \\
\hline
6. & 1 & \makecell[c]{0.5 \\ 0.5 \\ 0.7071} & 1.0000 & \makecell[c]{0.5000 \\ 0.5000 \\ 0.7071} & 2.4620e-6 & 0.0000 \\
\hline
7. & 1.7321 & \makecell[c]{0.5774 \\ 0.5774 \\ 0.5774} & 1.7321 & \makecell[c]{0.5774 \\ 0.5774 \\ 0.5774} & 2.4620e-6 & 0.0000 \\
\hline
8. & 1 & \makecell[c]{0.7071 \\ 0 \\ -0.7071} & 0 & \makecell[c]{NaN \\ NaN \\ NaN} & N/A & N/A \\
\hline
9. & 1 & \makecell[c]{0.7071 \\ 0 \\ -0.7071} & 1.0000 & \makecell[c]{0.7071 \\ 0 \\ -0.7071} & 3.1500e-5 & 0.0000 \\
\hline
\end{tabular}
\caption{\label{tab:simulated}Results for the input cases using the simulated strain and lattice rotation data. Note that zero noise is present for any of these cases.}
\end{table}

From Table \ref{tab:simulated}, we see that in all cases (bar number 8), the computed Burgers vector is virtually identical to the expected result. Both magnitude percentage error and angular error between the computed and expected results are negligible for the input parameters given. This is as expected for the "noise-free" case, where the strain and lattice rotation fields had been mathematically determined such that the expected Burgers vectors would be computed exactly. Obviously this is not the case for experimental data, where the program can only provide an estimate of the Burgers vector given the measured strains and lattice rotations.

In interesting error case (8) arises when the dislocation line intersects the closed loop that defines the Burgers circuit. Upon adjusting the Burgers circuit such that the $z$-axis coordinate limits are [-122.5 \si{nm}, 122.5 \si{nm}] (such that the cuboid was 50 voxels wide in the $z$-direction) this intersection no longer occurs and the Burgers vector is computed correctly. Therefore, care must be taken to ensure that the Burgers circuit limits are defined such that the intersection with the dislocation line is avoided, as well as ensuring that the dislocation line does indeed pass through the chosen Burgers circuit.

\begin{figure}
    \centering
    \includegraphics[width=0.95\textwidth]{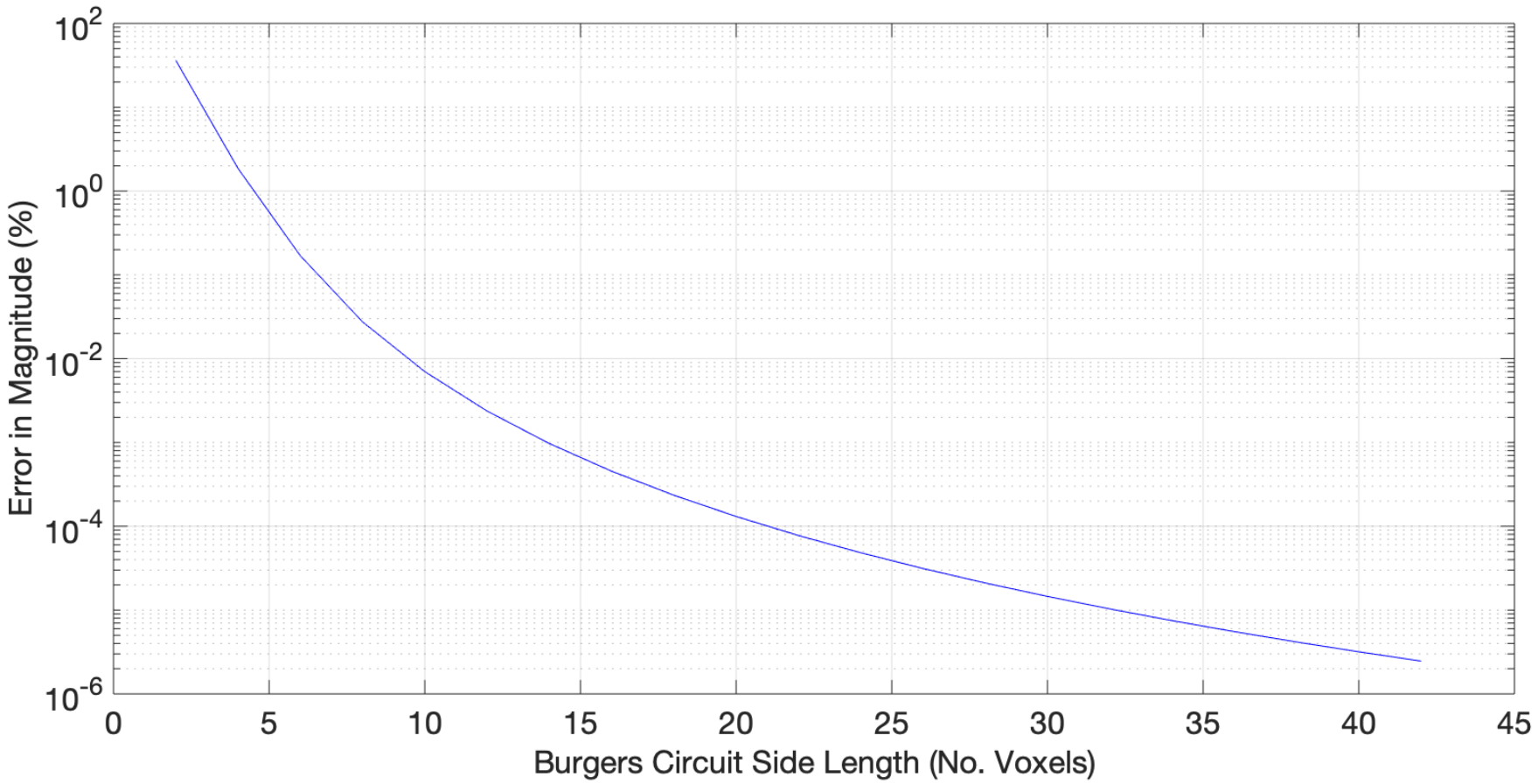}
    \caption{Test 1 (applied to Case 7) - Variation of computed Burgers vector with Burgers circuit size.  Angular error was found to be 0 for all sizes tested.}
    \label{fig:Test1}
\end{figure}

 Fig. \ref{fig:Test1} shows the results of Test 1 when applied to Case 7. It was seen that decreasing Burgers circuit size did not inherently increase angular error but did increase percentage error in magnitude. This effect was small for larger circuits but reached 1\% error at around 5 voxels wide and increased to 10\% at around 3 voxels wide. It should be noted that linear elasticity theory is valid at distances beyond a few Burgers vectors from the dislocation centre \cite{bacon}, and because voxel spacing is typically in the range of several nanometers we can assume that linear elasticity is valid for all voxels (unless the dislocation passes through the centre of one). Given this understanding, the reason for the deterioration in Burgers vector magnitude must be due to the numerical integration.

Reducing Burgers circuit size means that there are fewer voxels of data across which to integrate, and thus the integral is of a lower resolution. This will result in more substantial changes in strain and lattice rotation between adjacent voxels, leading to more significant error from numerical integration. This has a more severe impact for Burgers circuits that are small in size, as a single voxel has a proportionally greater contribution to the integral.

Additionally, and perhaps more importantly, smaller Burgers circuits will have their sides closer to the dislocation centre, and so the changes in value from voxel to voxel will be larger. Therefore the spatial gradients of the strains and lattice rotations at voxels on the Burgers circuit will tend to be steeper and are not properly captured when performing numerical integration. As a result, the error in the numerical integration will be greater in magnitude.

\begin{figure}
\centering
\subfloat[]{\includegraphics[width=0.445\textwidth]{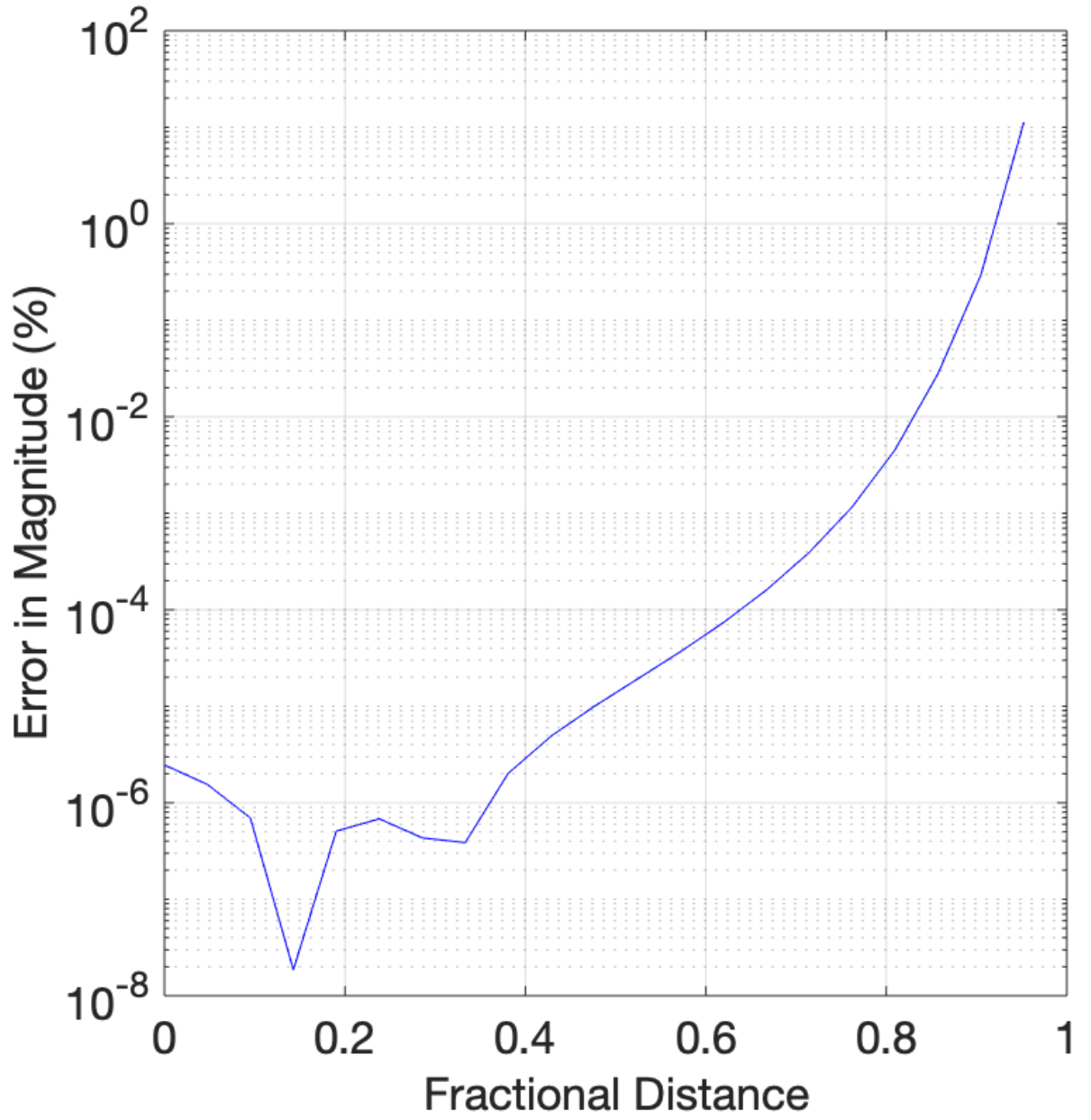}} \hspace{0.05\textwidth}
\subfloat[]{\includegraphics[width=0.445\textwidth]{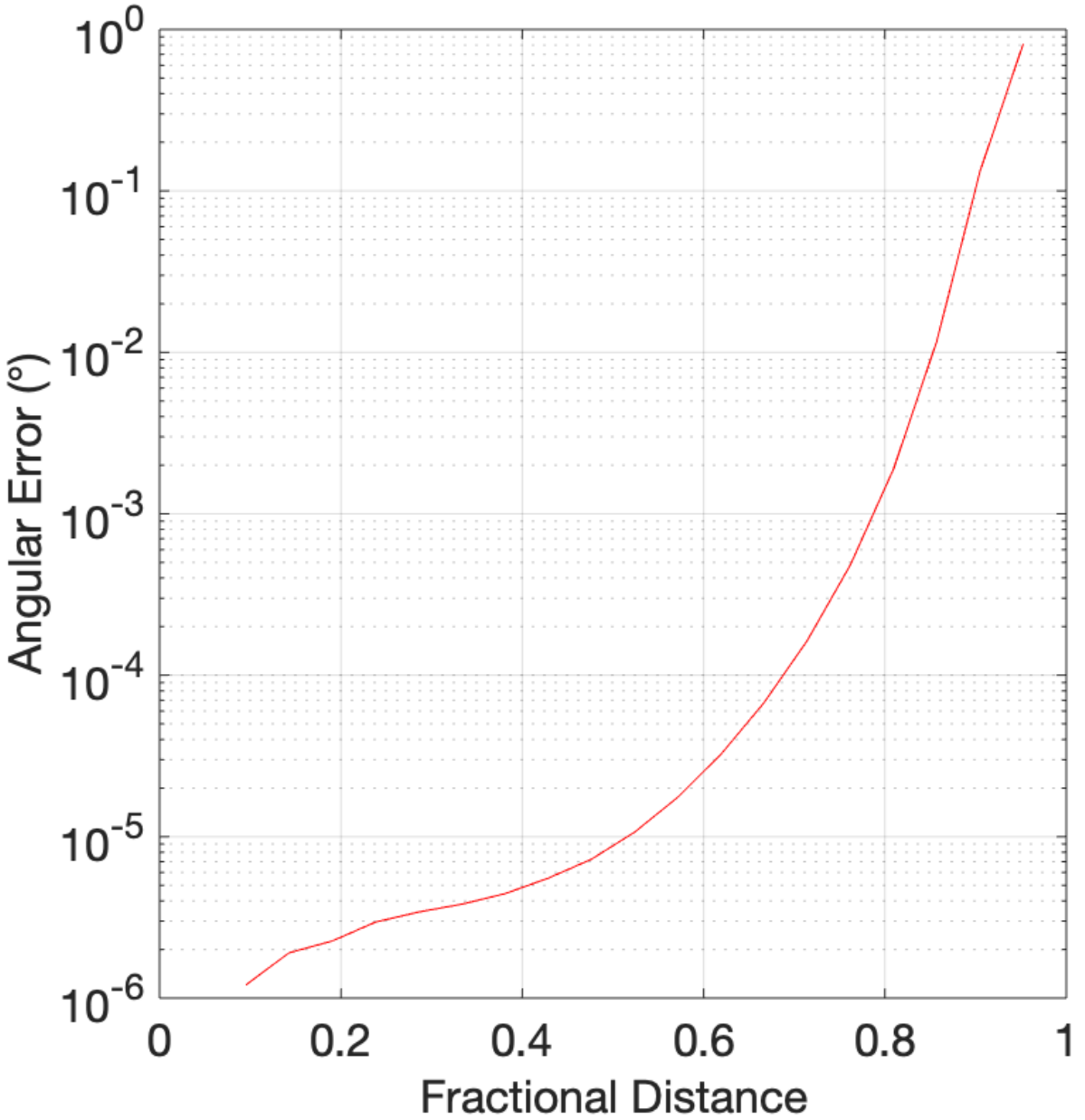}}
\caption{Test 2 (applied to Case 7) - Variation of computed Burgers vector with fractional distance of dislocation from Burgers circuit centre. Note the increase of both percentage error in magnitude (a) and angular error (b) as the ratio approaches 1.}
\label{fig:Test2}
\end{figure}

This effect is also evident when considering the case where the distance between the dislocation lines and the edge of a Burgers circuit with constant size is varied, such as when Test 2 was applied to Case 7 (see Fig. \ref{fig:Test2}). Here, despite the number of voxels contributing to the Burgers circuit (and therefore input data) staying the same, there was an increase in percentage magnitude error similar to that observed in Test 1 as the dislocation is moved closer to one side of the Burgers circuit. There was also an increase in angular error (reaching 0.1\degree at a ratio of around 0.9), which is likely due to an asymmetry in strain and lattice rotation magnitudes on opposite sides of the Burgers circuit. The side closer to the dislocation suffers from large changes in magnitude between adjacent voxels and therefore steeper spatial strain and lattice rotation gradients, resulting in greater error in the numerical integral. Meanwhile, the farther side benefits from smaller voxel-to-voxel changes in the strain and lattice rotation components. 

From these observations, it should be noted that, for an accurate computation of the Burgers vector, one must ensure that the dislocation is close to the centre of the Burgers circuit and that the Burgers circuit is of a sufficiently large size. The requirements will ultimately depend on the quality of the input data, as discussed in sections \ref{BCDI} and \ref{HRTKD}.

\begin{figure}
\centering
\subfloat[]{\includegraphics[width=0.445\textwidth]{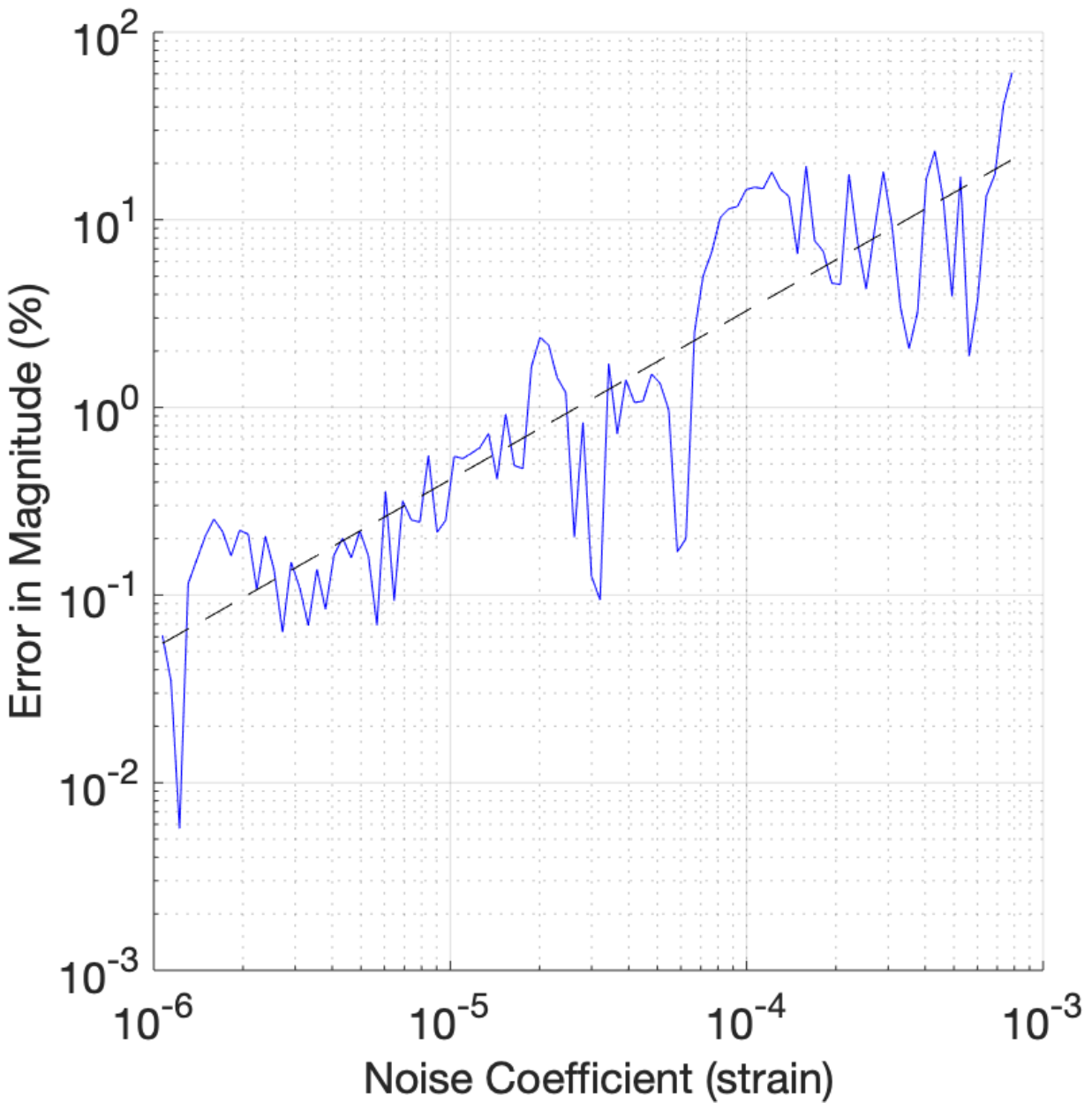}} \hspace{0.05\textwidth}
\subfloat[]{\includegraphics[width=0.445\textwidth]{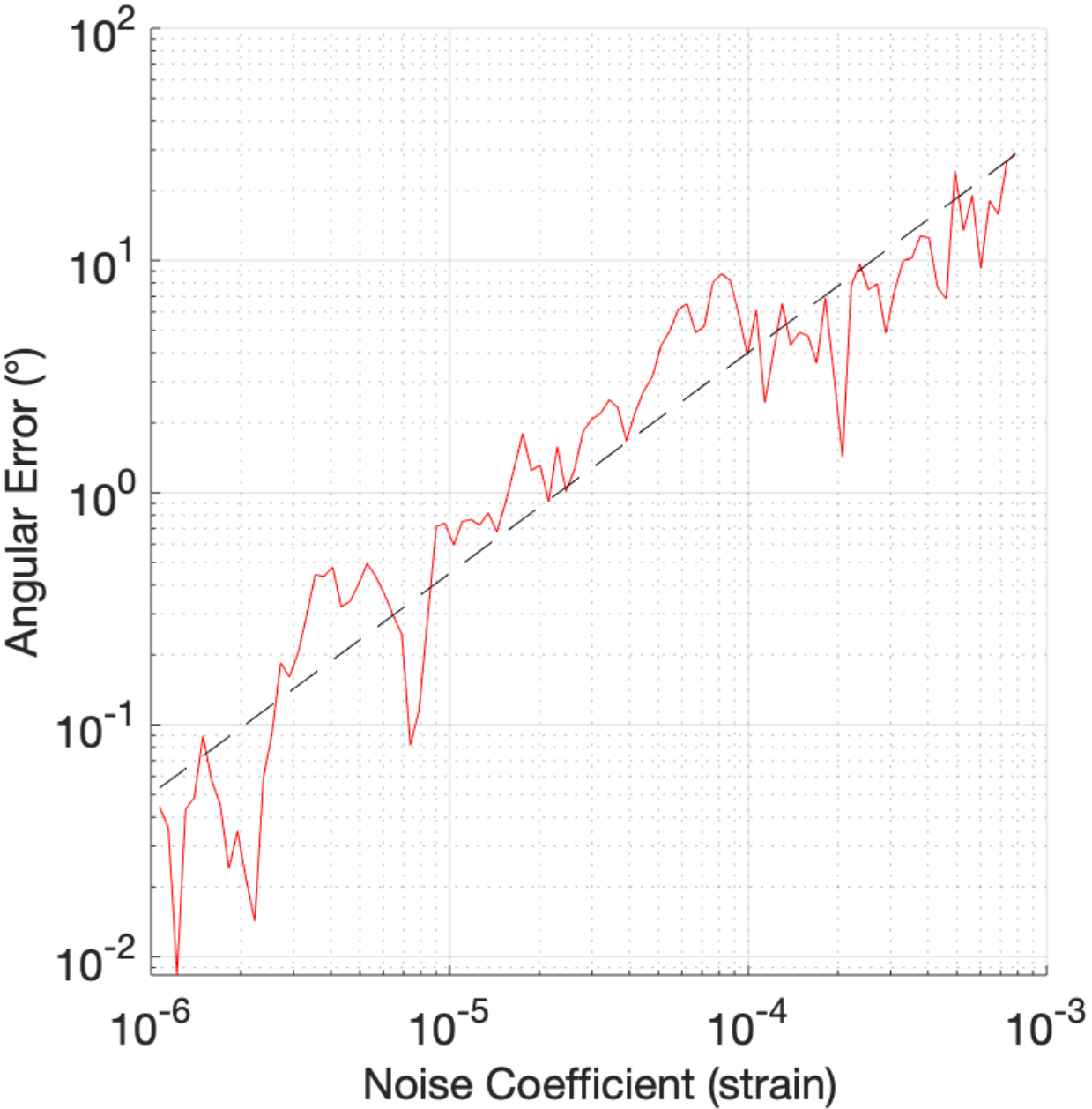}}
\caption{Test 3(A) (applied to Case 7) - Variation of computed Burgers vector with noise (single loop). Note the roughly linear general trend of increase of both percentage error in magnitude (a) and angular error (b) as the noise magnitude increases, with the line of best fit gradients being 0.927 and 0.971 respectively. The deviation of these gradients from 1 is a result of the inherent randomness of the noise.}
\label{fig:Test3A}
\end{figure}

\begin{figure}
\centering
\subfloat[]{\includegraphics[width=0.445\textwidth]{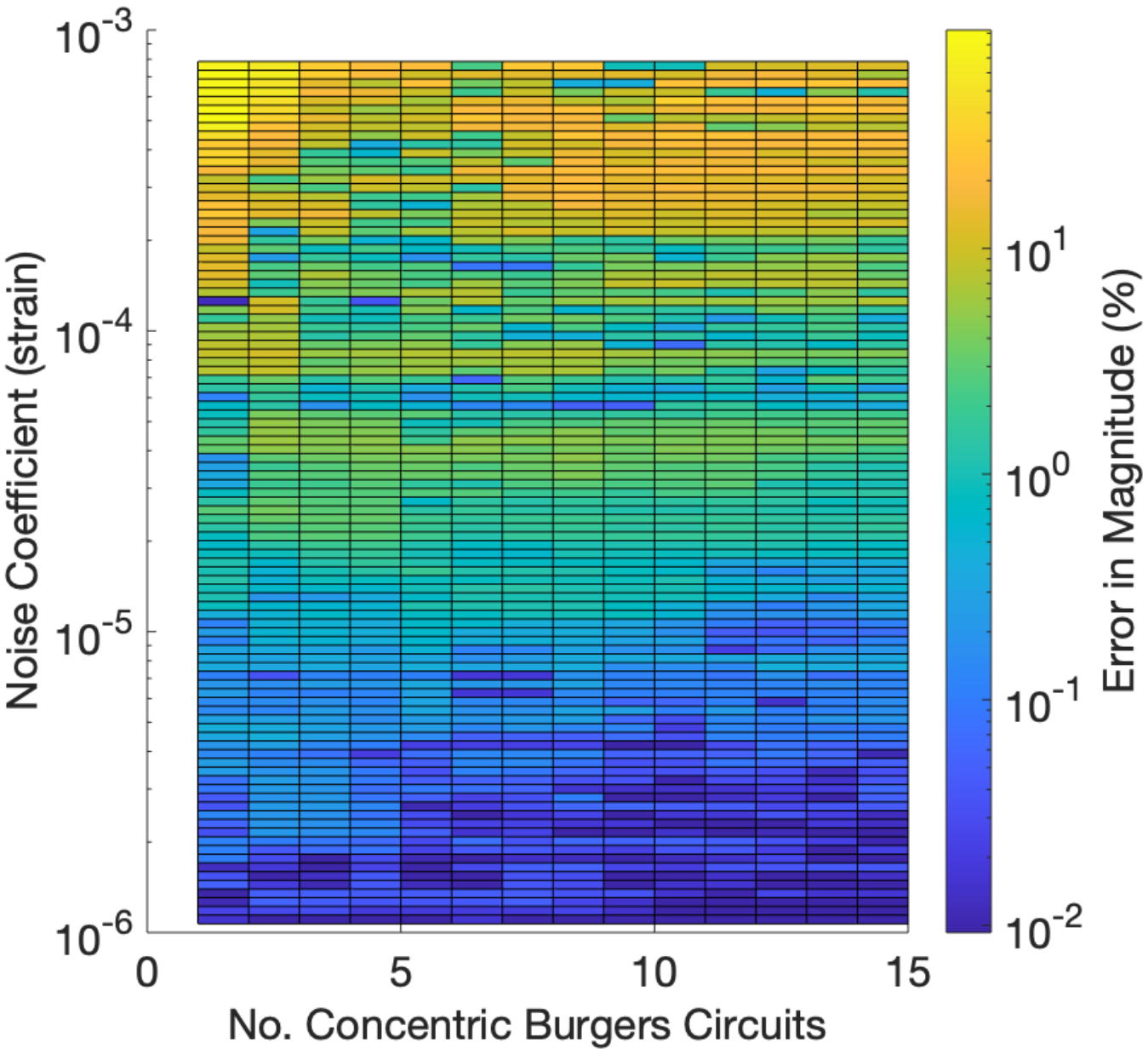}} \hspace{0.05\textwidth}
\subfloat[]{\includegraphics[width=0.445\textwidth]{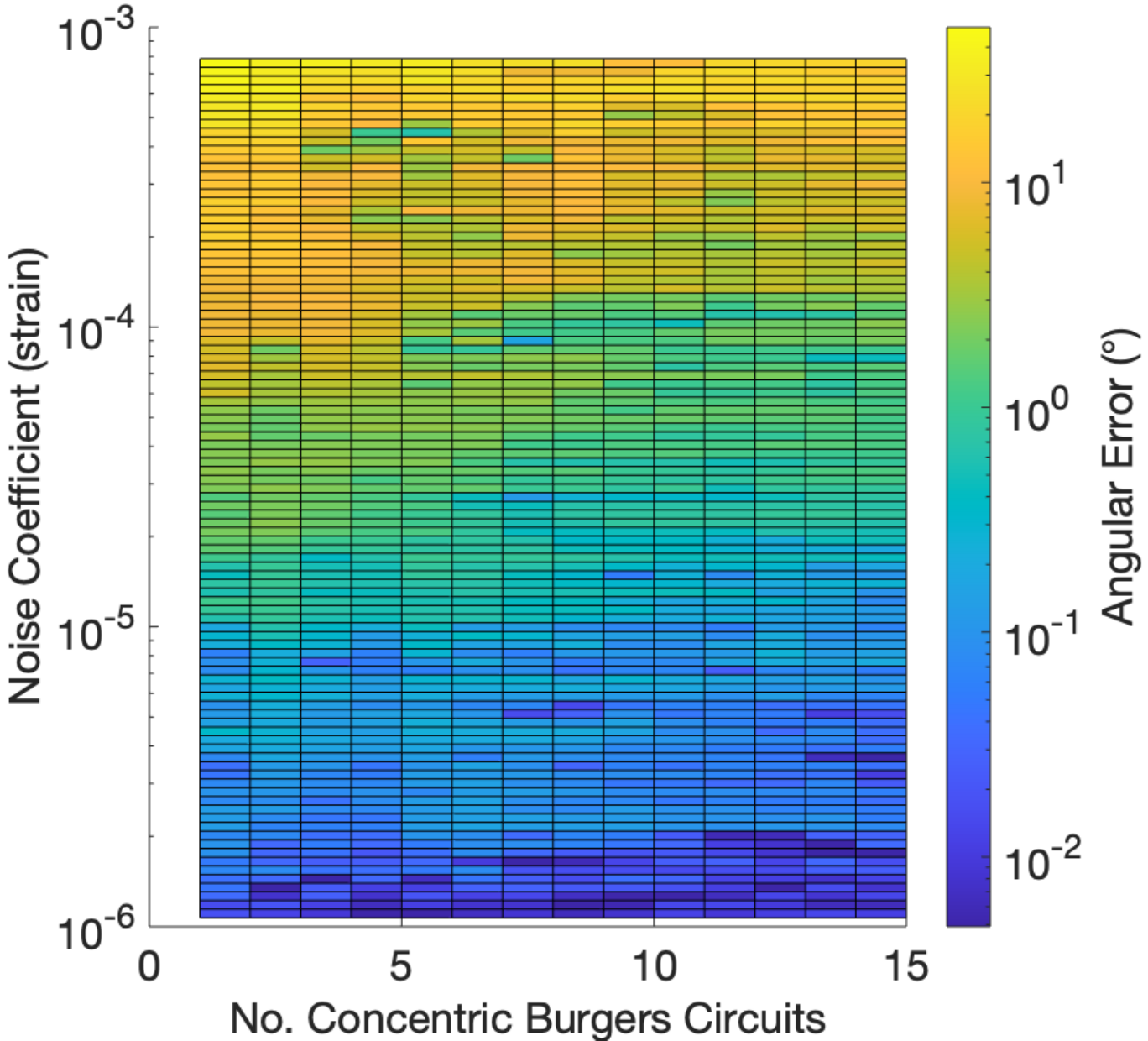}}
\caption{Test 3(B) (applied to Case 7) - Variation of computed Burgers vector with noise and no. concentric Burgers circuits. The percentage error in magnitude (a) and angular error (b) tend to increase as the noise magnitude increases and no. concentric Burgers circuits decreases.}
\label{fig:Test3B}
\end{figure}

Fig. \ref{fig:Test3A} shows the variation of computed Burgers vector with noise, with results produced by applying Test 3 to Case 7. It was found that increasing noise magnitude caused a roughly linear increase in both percentage error in magnitude and angular error, with the line of best fit gradients of the log-log graphs being 0.927 and 0.971 respectively. Note that the slight deviation of these gradients from 1 has resulted from the inherent randomness of the noise. At $\eta = 1\times10^{-4}$ there was roughly 3\% error in magnitude and 4\degree angular error  increasing to 6\% error in magnitude and 8\degree angular error at $\eta = 2\times10^{-4}$. These are realistic levels of uncertainty in experimental data (see sections \ref{BCDI} and \ref{HRTKD}). It should be noted that the simulated noise is defined entirely by the truncated Gaussian distribution, and the noise magnitude at one voxel is entirely independent of that in adjacent voxels.

As the noise magnitude approached $\eta = 1\times10^{-3}$ the computed Burgers vector deteriorated to the point that it had little relation to the expected result, likely as a result of the magnitudes of the strain field components becoming negligible compared to the noise itself. For very noisy data it can be beneficial to use smaller Burgers circuits in order to reduce the accumulation of noise in the numerical integration.

Fig. \ref{fig:Test3B} shows a general trend that increasing the number of concentric Burgers circuits decreased the errors caused by noise. The concentric Burgers circuits are all arranged such that they lie within surfaces $S_x$, $S_y$ and $S_z$ in Fig. \ref{fig:IntegrationLoop3D} b), and are as large as possible without crossing or overlapping at any points. The results obtained are as expected, as more computations of the Burgers vector allows for greater confidence that an individual result may be anomalous due to excessive noise (with $x$-, $y$- and $z$-components beyond a certain number of standard deviations from the mean values) and so can be discarded before averaging. Additionally, as the theoretical mean of the noise is zero, increasing the number of computed Burgers vectors to use for determination of a mean average Burgers vector should also generally result in a reduction in the effects of noise. However, due to the inherently random nature of the noise, these trends are never ideal and introducing another Burgers circuit that happens to be more significantly influenced by noise can actually worsen the final result. An example of this would be the fluctuation between high and low percentage error in magnitude as the number of concentric Burgers circuits is increased at the highest noise coefficient used in Test 3, i.e. the top row of elements in Fig. \ref{fig:Test3B} a).
%%%%%%%%%%%%%%%%%%%%%%%%%%%%%%%%%%%%%%

\section{Tests - BCDI Data} \label{BCDI}

Bragg coherent diffraction imaging (BCDI) allows the 3D-resolved experimental characterisation of morphology and lattice strain in microcrystals \cite{robinson,pfeifer}. Lattice reflections of a microcrystal illuminated with a coherent x-ray beam are used to measure coherent x-ray diffraction patterns (CXDPs), which are in turn used to determine the effective (complex) electron density in the sample. The phase of the complex electron density of a reflection in the $hkl$ direction, $\phi_{hkl}(\bm{r})$, is related to the displacement field of the crystal lattice structure, $\bm{u}(\bm{r})$, by the expression $\phi_{hkl}(\bm{r}) =  \bm{q}_{hkl}\cdot\bm{u}(\bm{r})$, where $\bm{q}_{hkl}$ is the reflection Bragg vector \cite{takagi}. By measuring at least three linearly-independent crystal reflections $\bm{u}(\bm{r})$ can be recovered, and the  displacement gradient $\bm{\beta}$ and thus the strain and lattice rotation fields calculated.

The data used for this test was taken from a micron-sized sample of high-purity tungsten containing a set of five dislocations and measured using BCDI. Six \{110\} crystal reflections were measured and their CXDPs recorded. Phase retrieval was then used to recover the complex-valued electron density for each reflection. This was projected back into a common, orthogonal sample coordinate frame with $5 \times 5 \times 5$ \si{nm}$^{3}$ voxel size. Spatial resolution was quantified by differentiating line profiles of electron density amplitude across the object-air interface and fitting these with a Gaussian profile. The average 3D spatial resolution, taken as $2\sigma$ of the fitted Gaussian, was $22$ \si{nm}. The data was processed to determine the strain and lattice rotations $\bm{\varepsilon}$ and $\bm{\omega}$. For further information on the manufacture of the material specimen, the BCDI measurements, the processing of the data and the strain and lattice rotation fields, refer to Hofmann et al. \cite{hofmann}.

\begin{figure}
\centering
\subfloat[]{\includegraphics[width=0.27\textwidth]{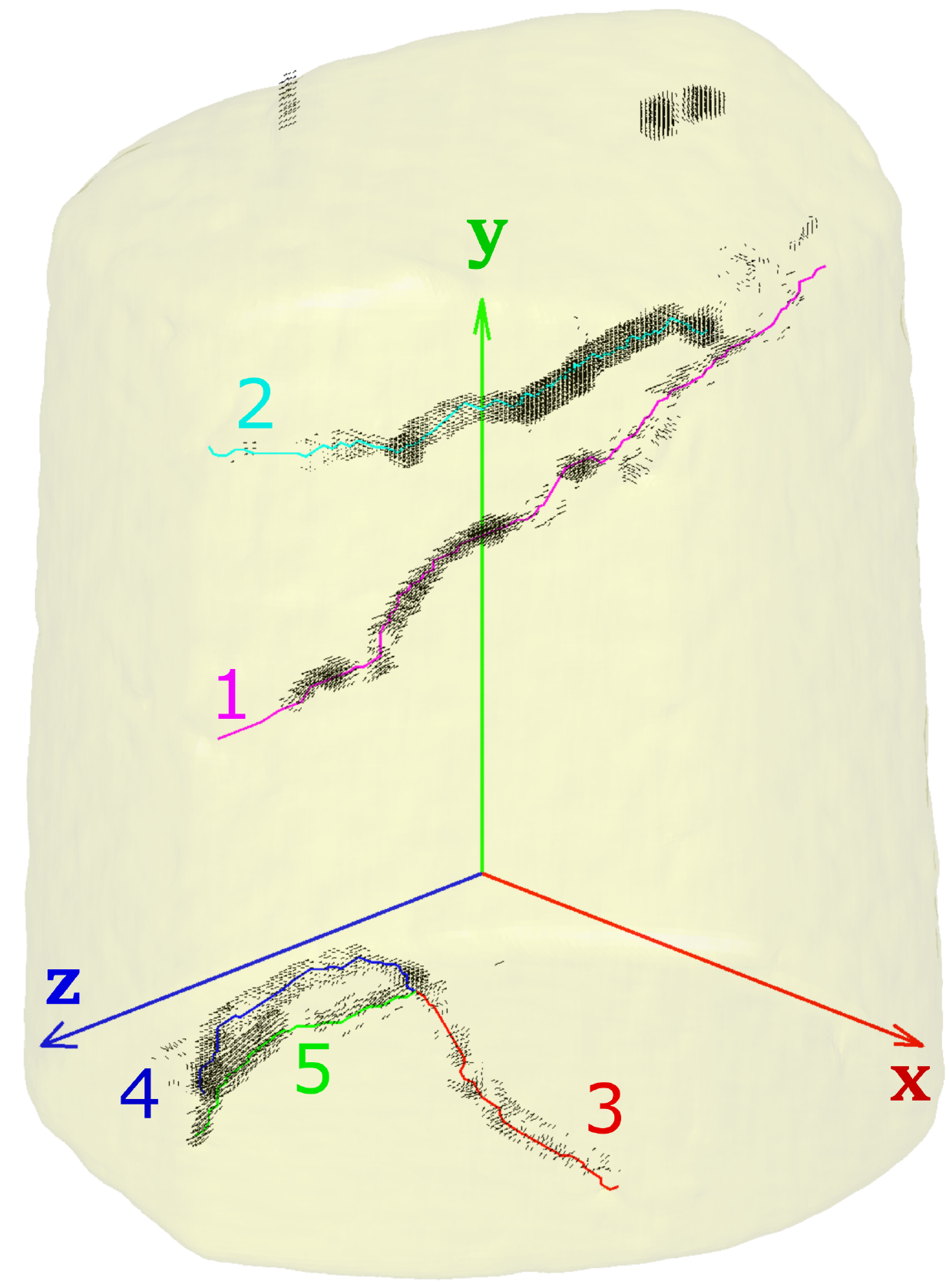}} \hspace{0.03\textwidth}
\subfloat[]{\includegraphics[width=0.33\textwidth]{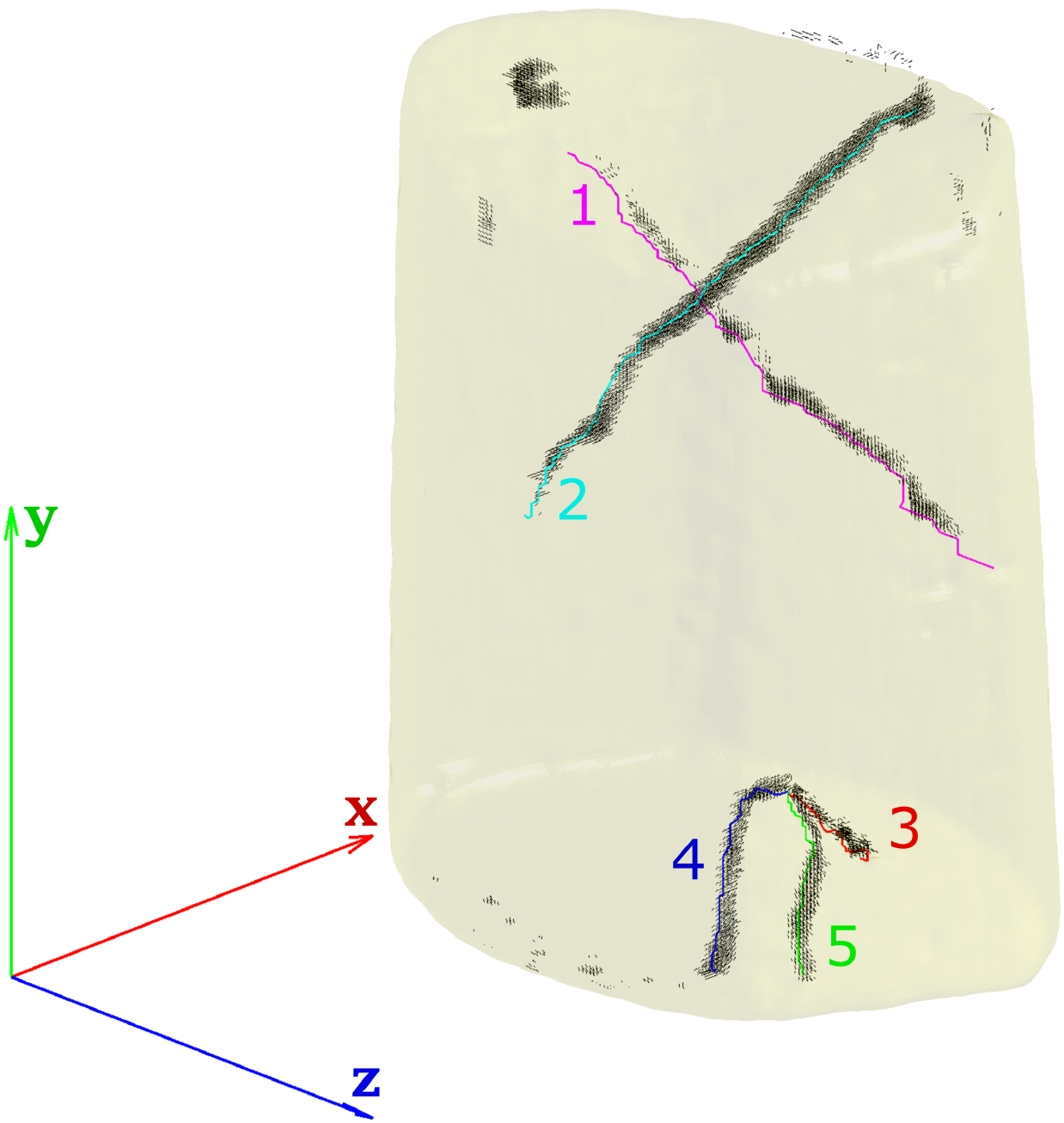}} \hspace{0.03\textwidth}
\subfloat[]{\includegraphics[width=0.3\textwidth]{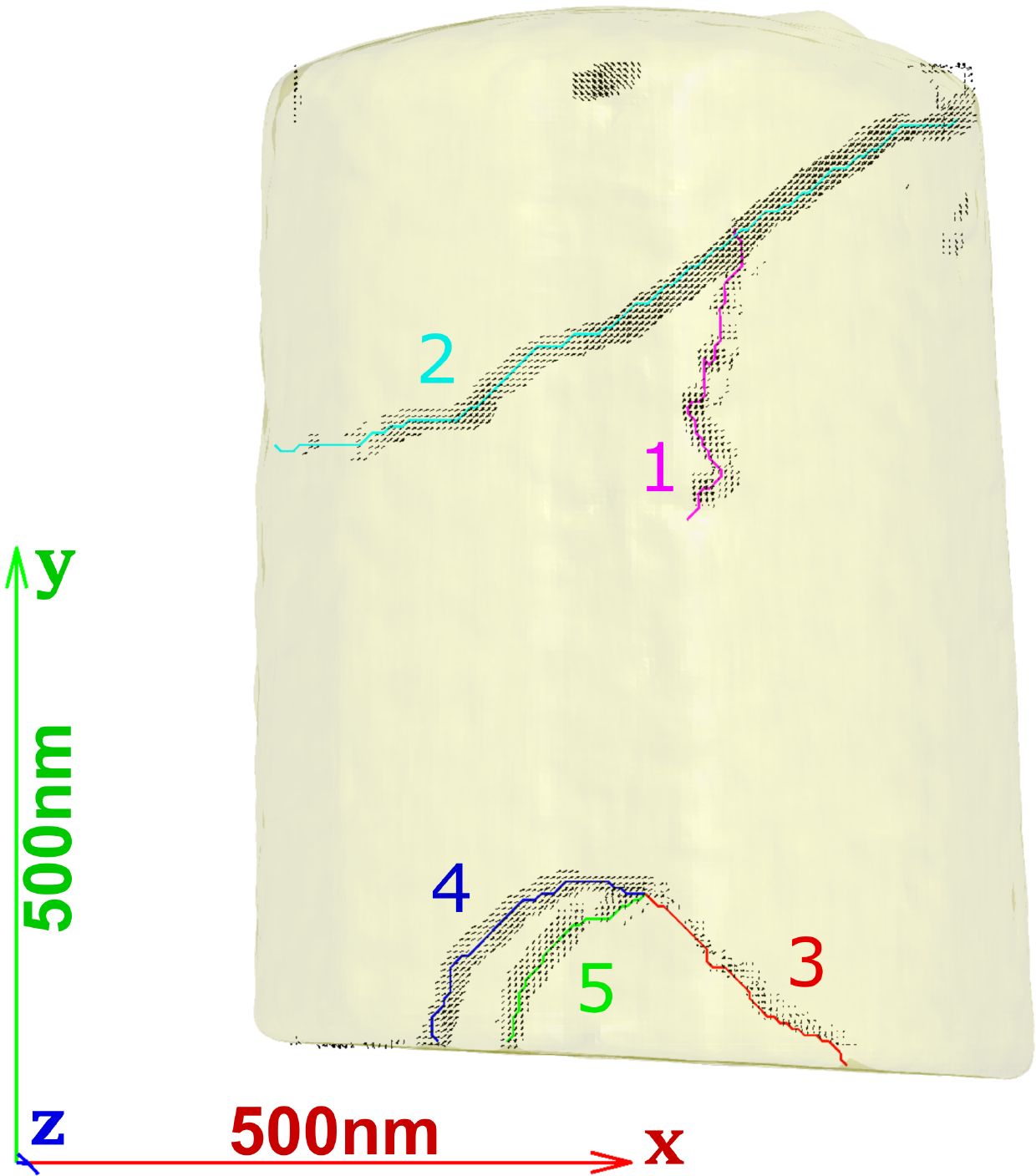}}
\caption{The 3D plot produced by rastering a 5x5x5 voxel Burgers circuit over the domain to build up a Burgers vector map, using the BCDI data as an input. The regions of black quivers display computed Burgers vectors above a cutoff magnitude. The material specimen shape and the experimentally-observed dislocation lines are also plotted. Numbers labelling each dislocation have been added and three views of the sample from different directions are shown. The coordinate axes are all 500 \si{nm} for reference scale. Note that the origin of the coordinate system is at the centre of the material specimen, not at the centre of the triad.}
\label{fig:BCDIplotter}
\end{figure}

The locations of the five dislocations were first determined and plotted using a standard Burgers circuit size of 5x5x5 voxels rastered over the 3D sample volume. Fig. \ref{fig:BCDIplotter} shows the specimen shape and known dislocations superposed on the field of quivers, with the quivers depicting the direction and magnitude of the computed Burgers vector at each point.

As can be seen in Fig. \ref{fig:BCDIplotter}, the approach was able to reliably identify the 3D path of dislocations in an automated fashion. Increasing the size of the standard Burgers circuit that iterates through the data resulted in more accurate Burgers vectors near the centres of the dislocation lines at the cost of thicker regions of computed Burgers vectors around each dislocation line. The reasoning for the latter effect is that because the Burgers circuit is larger, it catches dislocations from a greater distance away. Note that the size of 5x5x5 voxels was used to achieve thin regions of computed Burgers vectors at the expense of accuracy, and increasing the size to 9x9x9 would significantly reduce the magnitude and angular error.

The field of Burgers vectors is stored in one array, and it is straightforward to use the position of a computed Burgers vector in the array to determine the Cartesian coordinates of the centre of the Burgers circuit used to calculate it. Another (preferably larger) Burgers circuit that surrounds this point can then be manually chosen to allow for closer inspection of the Burgers vector.

Upon initial inspection on the resulting 3D plot, the dislocations were then examined more carefully. Large limits for the Burgers circuit were used at first and they were then tightened over a process of iteration to achieve a size of 9x9x9 voxels, which was still sufficient to produce accurate results. The final size used for dislocation 3 was actually 8x9x9 voxels to better fit the shape of the dislocation line.

\begin{samepage}
Chosen Burgers circuit coordinate limits (as depicted in Fig. \ref{fig:ExampleTestFig}) for each Dislocation (\si{nm}):
\begin{enumerate}
    \item[1.] x = [42.5,82.5], y = [82.5,42.5], z = [222.5,262.5]
    \item[2.] x = [-177.5,-137.5], y = [137.5,97.5], z = [-72.5,-32.5]
    \item[3.] x = [67.5,102.5], y = [-317.5,-357.5], z = [52.5,92.5]
    \item[4.] x = [-92.5,-52.5], y = [-242.5,-282.5], z = [82.5,122.5]
    \item[5.] x = [-67.5,-27.5], y = [-257.5,-297.5], z = [152.5,192.5]
\end{enumerate}
\end{samepage}

It should be noted that the chosen Burgers circuit limits are merely examples, and in fact there are a great many that can be chosen in order to achieve accurate computation of the Burgers vectors.

Strain and lattice rotation data was also extracted by regions unaffected by the presence of defects and the standard deviation (and therefore effectively the noise magnitude) was computed for each component of strain and lattice rotation.

\begin{table}
\begin{tabular}{||m{4.5em}|m{5.7em}|m{5.7em}|m{5.1em}|m{4.2em}|m{4.7em}||}
\hline
Dislocation & True Direction (Lattice Coordinates) & True Direction (Lab Coordinates)  & Computed Magnitude (\si{\angstrom}) & Computed Direction & Angular Error (\degree) \\
\hline
1 & $\frac{1}{2} [\overline{1}\overline{1}\overline{1}]$ & \makecell[c]{-0.2799 \\ -0.5126 \\ 0.8117} & 2.7065 & \makecell[c]{-0.2752 \\ -0.5195 \\ 0.8090} & 0.5027 \\
\hline
2 & $\frac{1}{2} [\overline{1}11]$ & \makecell[c]{0.7854 \\ 0.5804 \\ 0.2151} & 2.7143 & \makecell[c]{0.7841 \\ 0.5808 \\ 0.2187} & 0.2205 \\
\hline
3 & $[\overline{1}00]$ & \makecell[c]{0.4405 \\ 0.0587 \\ 0.8958} & 3.0959 & \makecell[c]{0.4406 \\ 0.0507 \\ 0.8962} & 0.4590 \\
\hline
4 & $\frac{1}{2} [\overline{1}11]$ & \makecell[c]{0.7854 \\ 0.5804 \\ 0.2151} & 2.7166 & \makecell[c]{0.7856 \\ 0.5787 \\ 0.2191} & 0.2493 \\
\hline
5 & $\frac{1}{2} [\overline{1}\overline{1}\overline{1}]$ & \makecell[c]{-0.2799 \\ -0.5126 \\ 0.8117} & 2.6977 & \makecell[c]{-0.2760 \\ -0.5150 \\ 0.8115} & 0.2626 \\
\hline
\end{tabular}
\caption{\label{tab:BCDI}Results for the five dislocations present in the microcrystal observed by BCDI.}
\end{table}

From Table \ref{tab:BCDI} it can be seen that the computed directions for Burgers vectors are close to the directions expected from $\bm{q}_{hkl}\cdot\bm{b}$ analysis \cite{hofmann}, with angular error being within half a degree for all five dislocations. The computed magnitudes are also very consistent, and by taking a mean (adjusted for relative Burgers vector magnitudes in crystal lattice coordinates), the results predict a lattice parameter of $a = 3.121$\si{\angstrom}. This is $1.38$\% below the standard value of $3.1652$\si{\angstrom} \cite{featherstone,bolef}. Using the simple assumption of a Gaussian distribution for the noise, the average standard deviation and therefore noise magnitude was found to be $\eta = 1.0396\times10^{-4}$ by sampling `empty' regions of the data set, which would suggest an error of 6\degree in Burgers vector direction and a 3\% error in Burgers vector magnitude, given the results of Fig. \ref{fig:Test3A}. Thus the Burgers vectors computed using BCDI data are well within the expected uncertainty.

%%%%%%%%%%%%%%%%%%%%%%%%%%%%%%%%%%%%%%

\section{Tests - HR-TKD Data} \label{HRTKD}

High (angular) resolution electron backscatter diffraction (HR-EBSD) allows the measurement of strains and lattice rotations at the nano-scale. A scanning electron microscope (SEM) is used to record electron backscatter diffraction (EBSD) patterns over an array of points to measure the crystallographic orientation at each point \cite{wilkinson1,britton}. A cross-correlation based approach using a reference diffraction pattern allows for a significant improvement in angular resolution, with small shifts of features in the patterns relative to the reference used to determine lattice rotations and distortions \cite{wilkinson1,wilkinson2,wilkinson3,wilkinson4,wilkinson5}. This approach is referred to as HR-EBSD.\\
The transmission Kikuchi diffraction (TKD) variation of the technique operates by detecting the Kikuchi pattern from the bottom surface of a thin foil, which allows for a significantly improved spatial resolution over the conventional EBSD method \cite{keller}.

However, this method allows only for the measurement of the 2D in-plane deviatoric strain tensor, and was taken as an average over a depth of half the foil thickness in the case of the test data used \cite{yu}. Through the assumption of plane stress the full strain tensor required for computation of the Burgers vector can be calculated \cite{hlt}, which is reasonable given the specimen thickness of 40 \si{nm}\cite{yu}.

Here we examine HR-TKD data that has been recently reported from a single dislocation in tungsten \cite{yu}. The material used was a high-purity tungsten sample electro-polished to electron transparency. The dislocation observed was assumed to be straight with sense normal to the foil surface. A Zeiss Merlin SEM with a Bruker eFlash detector was used to carry out the HR-TKD measurements, with a TKD pattern size of $800 \times 600$ pixels and a scanning step size of $3.9$ \si{nm}. Cross-correlation analysis of the Kikuchi patterns was then performed. The estimated spatial resolution, due to the finite size of the electron interaction volume, was $12$ \si{nm}. The data was processed to generate maps for the components of the strain and lattice rotation tensors. For further information on the manufacture of the material specimen, the HR-TKD measurements taken, the processing of the data and the strain and lattice rotation fields, refer to the associated paper by Yu et al. \cite{yu}. It should be noted that a grain boundary is present in the the top left of the data set, manifesting as a diagonal line of significant strain and lattice rotation.

As with the BCDI data, the position of the dislocation centre was first found by rastering a 7x7 pixel Burgers circuit over the domain to build up a Burgers vector map.

\begin{figure}
\centering
\subfloat[]{\includegraphics[width=0.34\textwidth]{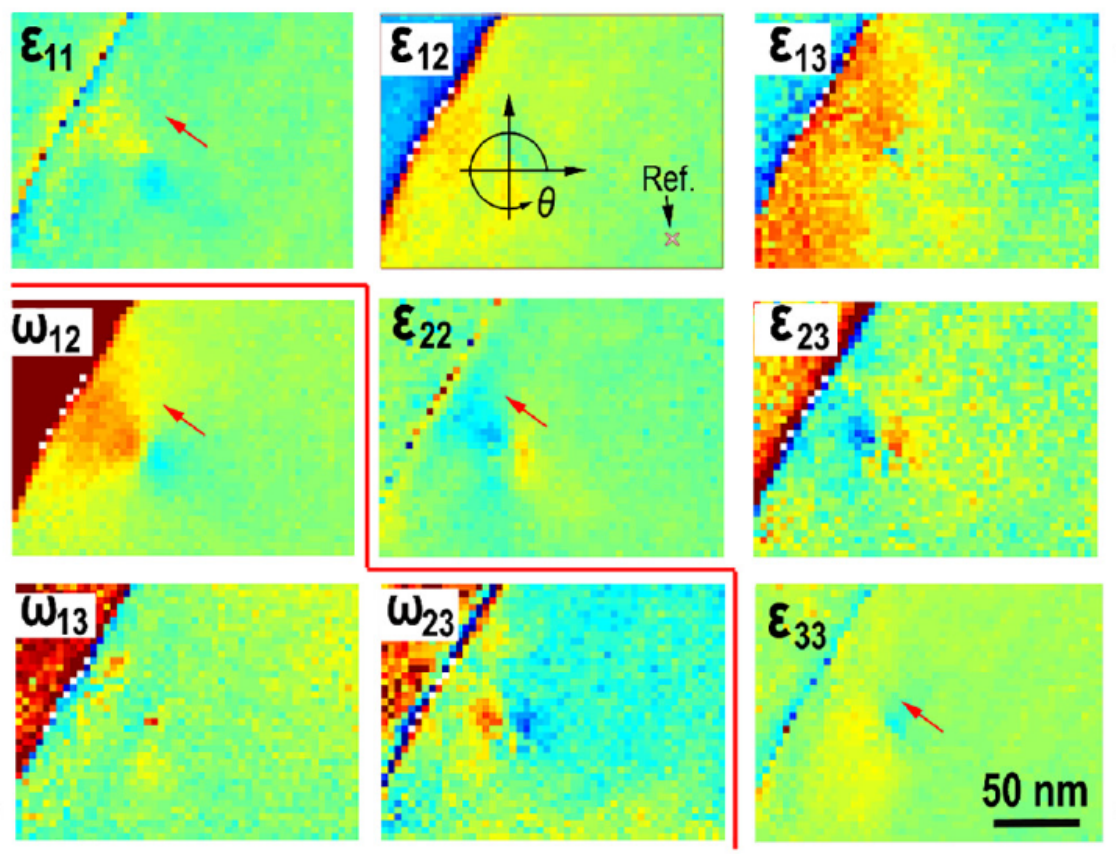}}
\subfloat[]{\includegraphics[width=0.64\textwidth]{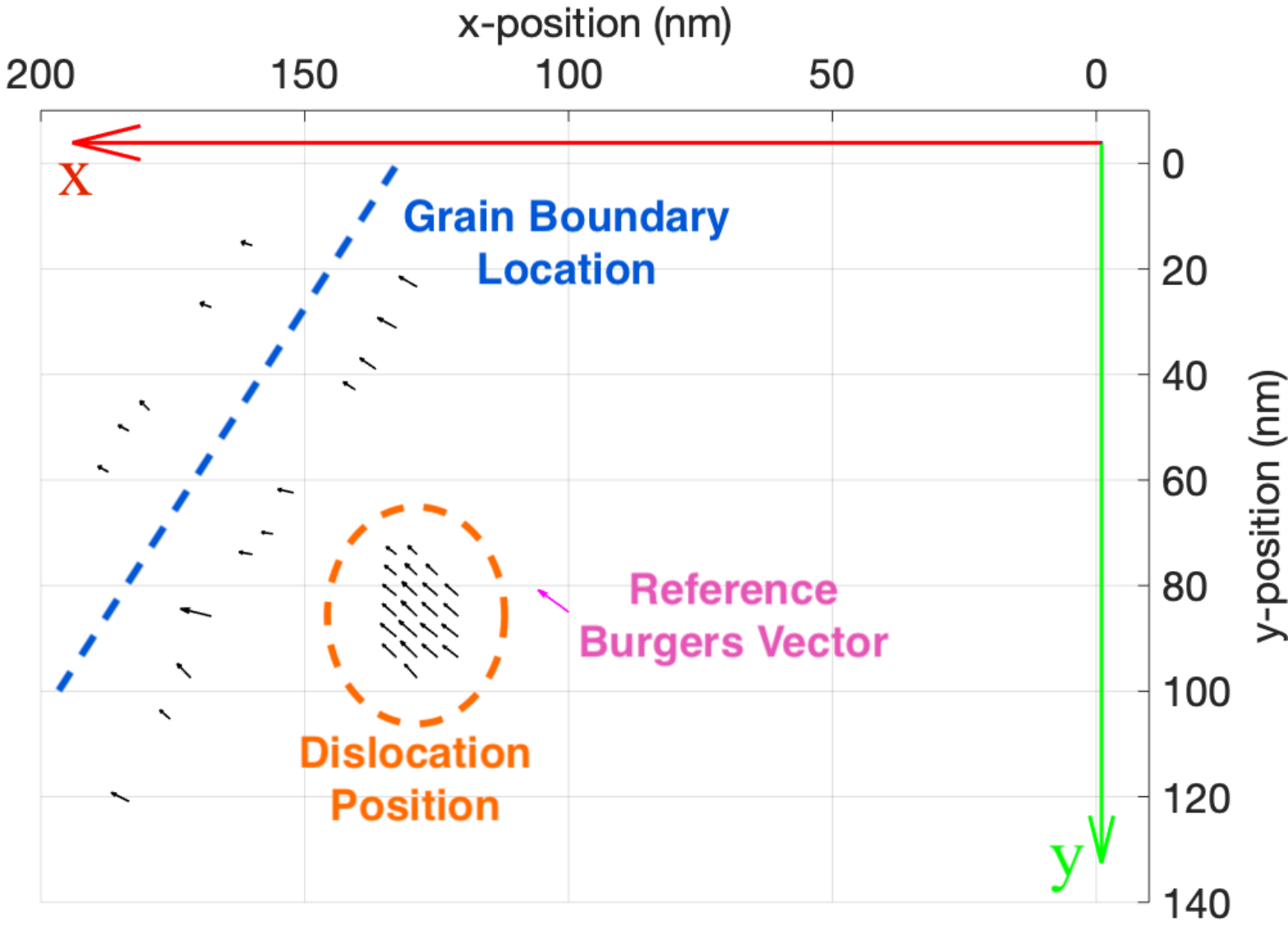}}
\caption{a) Full strain and lattice rotation tensor for the HR-TKD data (from Yu et al. \cite{yu}). b) 2D plot produced using the HR-TKD data as an input and then rastering a 7x7 pixel Burgers circuit over the domain to build up a Burgers vector map. The reference Burgers vector has also been added for comparison (magenta). Positions of the grain boundary and the dislocation are also annotated onto the plot.}
\label{fig:2D_Data_Plot}
\end{figure}

As depicted by Fig. \ref{fig:2D_Data_Plot}, our approach was able to discern the location of the dislocation as well as the grain boundary nearby. The region of black quivers identifies the position of the dislocation, as well as estimates for the direction of the computed Burgers vector. Also visible is a magenta quiver representing the expected Burgers vector direction from experimental measurement (and true magnitude from reference data). There is good agreement in direction and magnitude amongst the estimates, and fair agreement in direction between the estimates and the result expected by $\bm{g}\cdot\bm{b}$ contrast \cite{yu}.

Upon determination of the dislocation coordinates, a closer inspection was performed. The outermost Burgers circuit had the limits x = [20,40], y = [33,13] in nanometers. Several concentric Burgers circuits were used and the variation of percentage error in magnitude and angular error with Burgers circuit side length was recorded.

Strain and lattice rotation data was also extracted from regions unaffected by the presence of the dislocation or grain boundary and the standard deviation (and therefore effectively the noise magnitude) was computed for each component of strain and lattice rotation.

\begin{figure}
\centering
\subfloat[]{\includegraphics[width=0.445\textwidth]{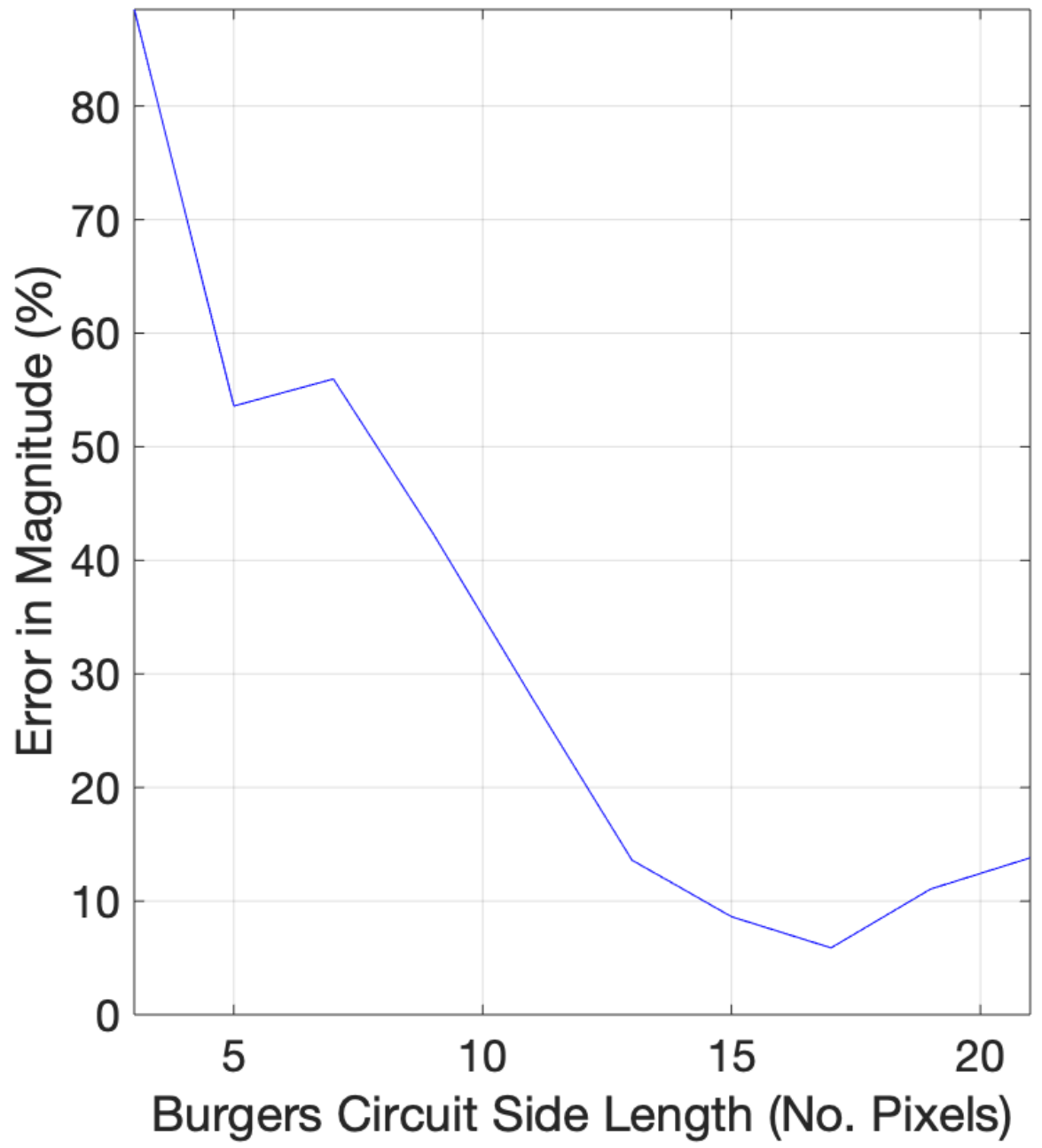}} 
\hspace{0.05\textwidth}
\subfloat[]{\includegraphics[width=0.445\textwidth]{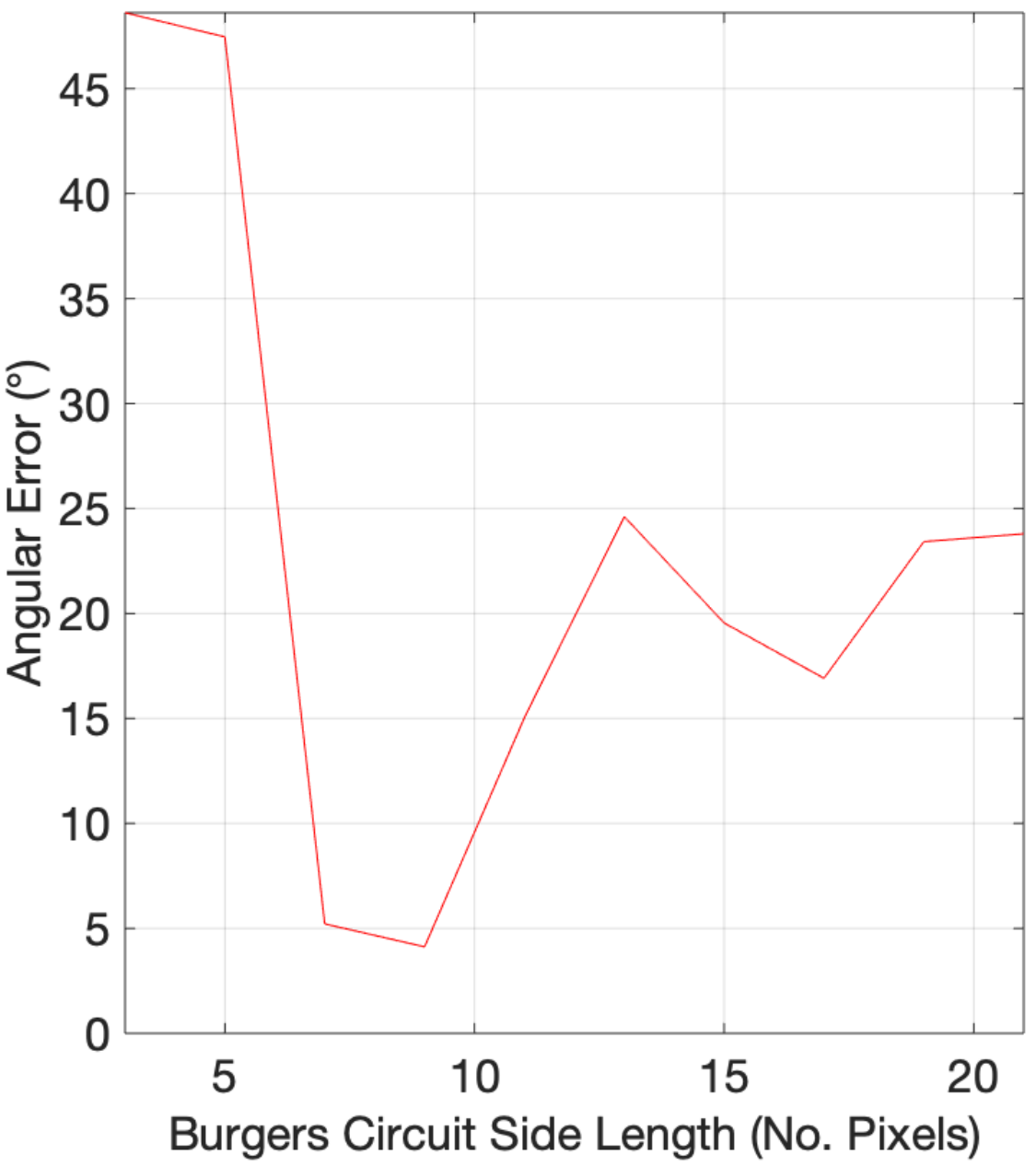}}
\caption{Variation of Computed Burgers Vector with Burgers Circuit Size (HR-TKD Data). The values of percentage error in magnitude (a) and angular error (b) at 17 pixels wide roughly match the values anticipated from Fig. \ref{fig:Test3A}.}
\label{fig:Test_2D_Data}
\end{figure}

Fig. \ref{fig:Test_2D_Data} shows that angular error tended to reduce with Burgers circuit side length to a minimum of 4.12\degree (with percentage error in Burgers vector magnitude of 42\% below the expected result of 2.7411\si{\angstrom}for tungsten \cite{featherstone,bolef}) at a Burgers circuit width of 9 pixels before increasing rapidly. There is also a local minimum of 16.9\degree angular error at 17 pixels. Meanwhile, the percentage error in magnitude was a minimum of 5.88\% at 17 pixels wide before increasing steadily as side length decreased (note that the magnitude of the Burgers vector decreased with decreasing side length). The significant percentage magnitude error for small Burgers circuit side length (for this data set, roughly 13 pixels and below) agrees with previously-discussed observations (see Fig. \ref{fig:Test1}). Assuming a Gaussian distribution for the noise, the average standard deviation and therefore noise magnitude was found to be $\eta = 2.7071\times10^{-4}$ by sampling `empty' regions of the data set, which is considerable. Observing Fig. \ref{fig:Test3A}, this would predict a percentage error of magnitude of at least 8\% and angular error of at least 12\degree. This roughly agrees with the results for the Burgers circuit of 17 pixels wide, an apparent "sweet-spot" in Burgers circuit size for this specific case.

Upon further observation of the original strain and lattice rotation data \cite{yu}, it was found that the grain boundary nearby had significant influence, especially on the $\varepsilon_{13}$ field. Increasing Burgers circuit size caused the integration loop to pass over the regions of intense strain and lattice rotation close to the grain boundary, almost crossing the grain boundary itself. Although theoretically this should not have affected Burgers vector calculation as the displacement field of the grain boundary was continuous across the path of integration, the discretisation of the data and numerical integration cause errors that worsen when the changes in magnitudes between adjacent pixels (i.e. the strain and lattice rotation gradients) increase. Therefore in this case it is not necessarily true that a larger Burgers circuit improves the result for computed Burgers vector, as illustrated by Fig. \ref{fig:Test_2D_Data}. The low angular error for smaller Burgers circuits of 7-9 pixels wide may support this also.

It was also noted that there was poor agreement of the measured $\varepsilon_{23}$ and $\omega_{23}$ components with simulated distortion fields for the known dislocation \cite{yu}.

Thus, even for this rather noisy experimental dataset, the approach presented in this paper can provide reasonable values for Burgers vector magnitude and direction. 

%%%%%%%%%%%%%%%%%%%%%%%%%%%%%%%%%%%%%%
\section{Conclusion}
We have presented an approach for determining dislocation Burgers vector magnitude and direction from experimentally measured lattice strain and rotation data in materials with low dislocation density. Applying this approach using simulated data worked flawlessly except for extreme cases where the Burgers circuit is very small or the dislocation line is very close to the integration loop. The addition of noise causes a degradation of the result that is proportional to the noise magnitude, though the computed Burgers vector still shares good agreement with the expected result with simulated noise at magnitudes typically found in experimental data. The use of concentric Burgers circuits can help to reduce errors in cases of very high noise.

For the BCDI data, the approach was able to reliably identify the 3D paths of all five dislocations in an automated fashion as well as accurately compute Burgers vectors with directions within half a degree of angular error to those expected (given appropriate Burgers circuit inputs). The computed magnitudes also showed good agreement with the standard value for the lattice parameter of tungsten.

The results using the HR-TKD data were also promising, as although deviation from the expected Burgers vector was much greater than for the BCDI data, there were a number of significant detrimental factors affecting computation and a result that still shows fair agreement was recovered. For this data also the rastering of a Burgers circuit across the sample space could be used to successfully identify the dislocation position.

Thus it has been demonstrated that the determination of Burgers vectors using strain and lattice rotation fields is a promising technique and is not only feasible for simulated data but also for a wide range of experimental data such as BCDI and even HR-TKD. It should be noted, however, that the resolution and noise of the data can have a significant impact on the results due to the numerical integration required for this technique. Furthermore, in order to study experimental data where the Burgers vector is unknown, clear and reliable bounds for the error in computed Burgers vector direction and magnitude will need to be developed. This will depend significantly on the data itself, in particular the presence of other crystal defects, experimental uncertainty, and the spatial resolution of the data. As a rudimentary estimate, we computed noise magnitude $\eta$ for the data sets investigated and used this value to predict the uncertainty in Burgers vector magnitude and direction.

\enlargethispage{20pt}

%\ethics{Insert ethics text here.}

\dataccess{The program created, data used and tests performed are freely available from the supplementary Github repository archived on Zenodo \cite{data}, preserved at the time of publication of this manuscript}.

\aucontribute{J. Cloete formulated, applied, and tested the approach reported in the paper, and wrote the manuscript. F. Hofmann and E. Tarleton designed the research project, provided supervision, and edited the manuscript. All authors have approved the final form of the paper.}

\competing{All authors declare there are no competing interests.}

\funding{The research was supported by the Engineering Undergraduate Research Opportunities Programme (EUROP), hosted by the Department of Engineering Science, University of Oxford.
F. Hofmann acknowledges funding from the European Research Council (ERC) under the European Union's Horizon 2020 research and innovation programme (Grant Agreement No. 714697).}

\ack{We thank Hongbing Yu for provision of the HR-EBSD data, and the Department of Engineering Science, University of Oxford for hosting this project.}

\appendix
\section{Mathematical Models of Singular Dislocations} \label{derivations}

It should be noted that the mathematical dislocation model used in this paper is singular; meaning that the discontinuity associated with the dislocation is confined to an infinitesimal line (that being the dislocation line). Note also that the displacement gradient  $\bm{\beta}$ is given by
\begin{align}
\bm{\beta} = \bm{\nabla u} = \begin{bmatrix}
\pdv{u_x}{x} & \pdv{u_x}{y} & \pdv{u_x}{z} \\
\pdv{u_y}{x} & \pdv{u_y}{y} & \pdv{u_y}{z} \\
\pdv{u_z}{x} & \pdv{u_z}{y} & \pdv{u_z}{z}
\end{bmatrix}
\end{align}

\subsection{Infinite Straight Edge Dislocation} \label{edge}

Let $x$, $y$ and $z$ be the three dimensions in a Cartesian coordinate system, with respective unit basis vectors $\bm{e_x}$, $\bm{e_y}$ and $\bm{e_z}$. The constant $v$ is the hypothetical Poisson's ratio of the material, and $b$ is the magnitude of the Burgers vector.

Consider an infinite straight edge dislocation, with sense $\bm{\xi}$ = $\mathbf{e_{z}}$
and Burgers vector $\mathbf{b}$ = $b\mathbf{e_{x}}$. The discontinuity of the dislocation is present at $x = 0$.

Anderson et al. \cite{anderson} provide an apt derivation of the components of displacement field $\bm{u}$, which are found to be
\begin{align}
u_{x} = \frac{b}{2\pi}\left[\tan^{-1}\left(\frac{y}{x}\right) + \frac{xy}{2(1-v)(x^2 + y^2)}\right], \\
u_{y} = -\frac{b}{2\pi}\left[\frac{1 - 2v}{4(1-v)}\ln\left(x^2 + y^2\right) + \frac{x^2 - y^2}{4(1-v)(x^2 + y^2)}\right]. 
\end{align}
Note that for this edge dislocation $u_{z} = 0$.

To find the elastic displacement gradient $\bm{\beta}$, the gradient of the displacement field, the spatial partial derivatives are taken:
\begin{align}
\beta_{11} = \pdv{u_x}{x} = -A\frac{y}{(x^2 + y^2)^2}\left[(3-2v)x^2 + (1-2v)y^2\right], \\
\beta_{12} = \pdv{u_x}{y} = A\frac{x}{(x^2 + y^2)^2}\left[(3-2v)x^2 + (1-2v)y^2\right], \\
\beta_{21} = \pdv{u_y}{x} = -A\frac{x}{(x^2 + y^2)^2}\left[(1-2v)x^2 + (3-2v)y^2\right], \\
\beta_{22} = \pdv{u_y}{y} = A\frac{y}{(x^2 + y^2)^2}\left[(1+2v)x^2 - (1-2v)y^2\right], \\
\beta_{13} = \beta_{23} = \beta_{31} = \beta_{32} = \beta_{33} = 0
\end{align}
where
\begin{align}
A = \frac{b}{4\pi(1-v)}
\end{align}

With the elastic displacement gradient thus represented by
\begin{align}
\bm{\beta}_{edge} = \begin{bmatrix}
\beta_{11} & \beta_{12} & 0 \\
\beta_{21} & \beta_{22} & 0 \\
0 & 0 & 0
\end{bmatrix}
\end{align}

\subsection{Infinite Straight Screw Dislocation} \label{screw}

Consider an infinite straight screw dislocation, with sense $\bm{\xi}$ = $\mathbf{e_{z}}$
and Burgers vector $\mathbf{b}$ = $b\mathbf{e_{z}}$ (note the dislocation is therefore right-handed). Let the plane of slip be confined to the positive x-axis (this is an important clarification for when we combine our results to produce a mixed dislocation).

Assuming that $u_{z}$ increases uniformly with angle $\theta$ about the dislocation core measured from the positive x-axis, the components of the elastic displacement $\bm{u}$ can be found by inspection \cite{bacon};
\begin{align}
u_x = 0; \quad u_y = 0 \\
u_z(r,\theta) = b\frac{\theta}{2\pi} \quad [0< \theta \leq2\pi]
\end{align}

Using the geometric relation between $x$, $y$ and $\theta$, the third expression can be re-written as
\begin{align}
u_z(x,y) = \begin{cases}
\frac{b}{2\pi}\tan^{-1}\left(\frac{y}{x}\right) & [0< \theta \leq \frac{\pi}{2}] \\
\frac{b}{2\pi}\left[\pi + \tan^{-1}\left(\frac{y}{x}\right)\right] & [\frac{\pi}{2}< \theta \leq \frac{3\pi}{2}] \\
\frac{b}{2\pi}\left[2\pi + \tan^{-1}\left(\frac{y}{x}\right)\right] & [\frac{3\pi}{2}< \theta \leq 2\pi]
\end{cases}
\end{align}

These all have the same partial derivatives, which are found to be
\begin{align}
\pdv{u_z}{x} = \beta_{31} = -\frac{b}{2\pi}\frac{y}{(x^2 + y^2)}, \\
\pdv{u_z}{y} = \beta_{32} = \frac{b}{2\pi}\frac{x}{(x^2 + y^2)}
\end{align}

To find the elastic displacement gradient $\bm{\beta}$, note that it will be the case that
\begin{align}
\beta_{11} = \beta_{12} = \beta_{13} = \beta_{21} = \beta_{22} = \beta_{23} = \beta_{33} = 0
\end{align}
and thus
\begin{align}
\bm{\beta}_{screw} = \begin{bmatrix}
0 & 0 & 0 \\
0 & 0 & 0 \\
\beta_{31} & \beta_{32} & 0
\end{bmatrix}
\end{align}

\subsection{Infinite Straight Mixed Dislocation} \label{mixed}
A mixed dislocation with sense $\bm{\xi}$ = $\mathbf{e_{z}}$ and Burgers vector $\bm{b}$ constrained to the $x-z$ plane can be described by linear combination of the previously-described edge and screw cases. Projection of the Burgers vector onto the z-axis and x-axis gives $\bm{b}_{screw} = \bm{b}\cos(\alpha)$ and $\bm{b}_{edge} = \bm{b}\sin(\alpha)$ respectively, where $\alpha$ is the angle subtended from the z-axis to the Burgers vector, such that $\alpha = 0\degree$ corresponds to a screw dislocation and $\alpha = 90\degree$ to an edge dislocation. The displacement gradients of the two cases can be superposed because they are independent of each other in linear isotropic elasticity \cite{bacon}. It follows, therefore, that the displacement gradient of the mixed dislocation is given by
\begin{align}
    \bm{\beta} = \bm{\beta}_{edge}\sin(\alpha) + \bm{\beta}_{screw}\cos(\alpha)
\end{align}
where $\bm{\beta}_{edge}$ and $\bm{\beta}_{screw}$ are as described in the previous sections.

The result is the following tensor:
\begin{align}
\bm{\beta} = \begin{bmatrix}
\beta_{11} & \beta_{12} & 0 \\
\beta_{21} & \beta_{22} & 0 \\
\beta_{31} & \beta_{32} & 0
\end{bmatrix}
\end{align}
where
\begin{align}
\beta_{11} = -\frac{b \sin(\alpha)}{4\pi(1-v)}\frac{y}{(x^2 + y^2)^2}\left[(3-2v)x^2 + (1-2v)y^2\right], \\
\beta_{12} = \frac{b \sin(\alpha)}{4\pi(1-v)}\frac{x}{(x^2 + y^2)^2}\left[(3-2v)x^2 + (1-2v)y^2\right], \\
\beta_{21} = -\frac{b \sin(\alpha)}{4\pi(1-v)}\frac{x}{(x^2 + y^2)^2}\left[(1-2v)x^2 + (3-2v)y^2\right], \\
\beta_{22} = \frac{b \sin(\alpha)}{4\pi(1-v)}\frac{y}{(x^2 + y^2)^2}\left[(1+2v)x^2 - (1-2v)y^2\right], \\
\beta_{31} = -\frac{b \cos(\alpha)}{2\pi}\frac{y}{(x^2 + y^2)}, \\
\beta_{32} = \frac{b \cos(\alpha)}{2\pi}\frac{x}{(x^2 + y^2)}.
\end{align}
Note that we can convert the result into strain and lattice rotation tensors using the definitions $\bm{\varepsilon} = \frac{1}{2} \left(\bm{\beta} + \bm{\beta}^{T}\right)$ and $\bm{\omega} = \frac{1}{2} \left(\bm{\beta} - \bm{\beta}^{T}\right)$ \cite{hlt,rima}.

\subsection{Rotating the Mixed Dislocation} \label{rotating}

To rotate the mixed dislocation such that both $\bm{\xi}$ and $\bm{b}$ can take any orientation, it was found that applying 3 consecutive rotations about coordinate axes performs this task effectively. Each of the rotations causes a rotation of the entire displacement gradient field, however they are applied in an order such that the following procedure is effectively carried out:
\begin{enumerate}
    \item[1.] Rotate $\bm{b}$ anticlockwise by angle $\psi$ about the z-axis. This, paired with the aforementioned use of angle $\alpha$, allows $\bm{b}$ to take any orientation relative to $\bm{\xi}$. Note $\bm{\xi}$ is unaffected.
    \item[2.] Rotate $\bm{\xi}$ anticlockwise by angle $\theta$ about the y-axis. This allows $\bm{\xi}$ to take any orientation within the $x-z$ plane. Note both $\bm{\xi}$ and $\bm{b}$ are affected equally.
    \item[3.] Rotate $\bm{\xi}$ anticlockwise by angle $\phi$ about the z-axis. Note both $\bm{\xi}$ and $\bm{b}$ are affected equally. This, paired with the previous rotation, allows $\bm{\xi}$ to take any orientation.
\end{enumerate}

\begin{figure}
\centering
\subfloat[]{\includegraphics[width=0.445\textwidth]{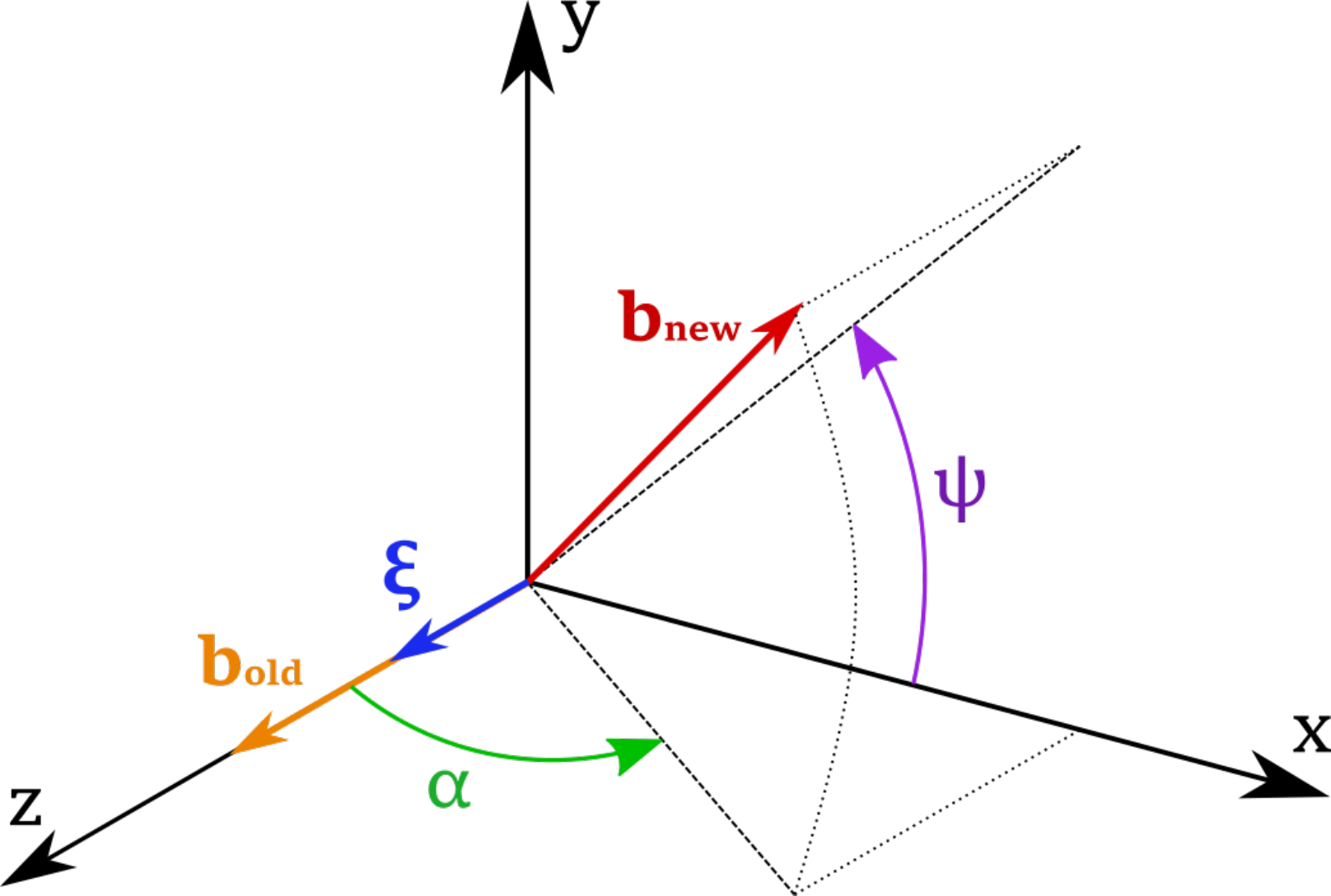}} \hspace{0.05\textwidth}
\subfloat[]{\includegraphics[width=0.445\textwidth]{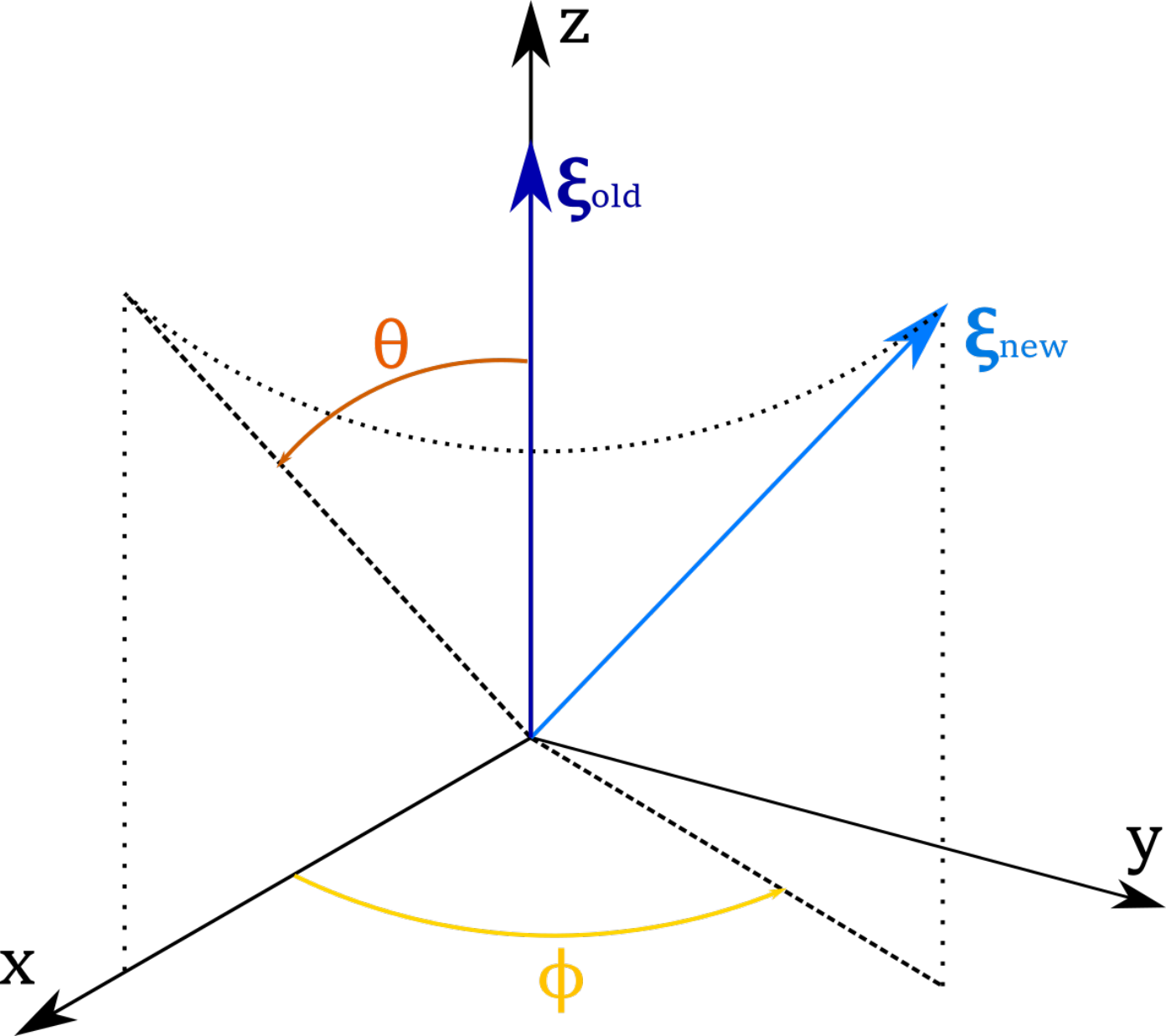}}
\caption{Summary of the rotation of the mixed dislocation. a) Setting the orientation of $\bm{b}$ relative to $\bm{\xi}$. b) Setting the orientation of $\bm{\xi}$.}
\label{fig:rotations}
\end{figure}

The method is summarised visually in Figure \ref{fig:rotations}.
The three rotations can each be described separately by rotation matrices, which can be combined to form an overall rotation matrix:

\begin{align}
\bm{R}_{1} = 
\begin{bmatrix}
\cos(\psi) & -\sin(\psi) & 0 \\
\sin(\psi) & \cos(\psi) & 0 \\
0 & 0 & 1
\end{bmatrix}, \\
\bm{R}_{2} =
\begin{bmatrix}
\cos(\theta) & 0 & \sin(\theta) \\
0 & 1 & 0 \\
-\sin(\theta) & 0 & \cos(\theta)
\end{bmatrix}, \\
\bm{R}_{3} =
\begin{bmatrix}
\cos(\phi) & -\sin(\phi) & 0 \\
\sin(\phi) & \cos(\phi) & 0 \\
0 & 0 & 1
\end{bmatrix}, \\
\bm{R} = \bm{R}_{3} \bm{R}_{2} \bm{R}_{1}
\end{align}
$\bm{R}$ rotates a tensor expressed in the dislocation frame $(x',y',z')$ to the global frame $(x,y,z)$.  The tensor field $\bm{\beta}(x,y,z)$ of a dislocation at arbitrary orientation, is then found by rotating the field expressed in the dislocation frame $\bm{\beta}'(x',y',z')$;
\begin{align}
\bm{\beta}(x,y,z) = \bm{R}\bm{\beta}'\left(x',y',z' \right)\bm{R}^{T}
\end{align}

% For the tensor field $\bm{\beta}(x,y,z)$, the rotated field $\bm{\beta}_{R}(x,y,z)$ is given by
% \begin{align}
% \bm{\beta}_{R}(x,y,z) = \bm{R}\bm{\beta}\left(x',y',z' \right)\bm{R}^{T}
% \end{align}

The variables $x'$, $y'$ and $z'$ are the field point coordinates in the rotated dislocation frame and are found by rotating the field points in the global frame $x$, $y$, $z$ in the opposite direction;
\begin{align}
\begin{bmatrix}
x' \\
y' \\
z'
\end{bmatrix}
= \bm{R}^{T}
\begin{bmatrix}
x \\
y \\
z
\end{bmatrix}
\end{align}
%
% The change of input arguments is required because we must first transform the coordinate system in which the field $\bm{\beta}$ was originally defined such that, upon applying the rotation to $\bm{\beta}$ itself, the rotated field $\bm{\beta}_{R}$ is defined in the original, untransformed coordinate system ($x$, $y$ and $z$).
%
Thus the required expressions for an infinite straight mixed dislocation of any orientation have been derived.

\section{The Nye Tensor Approach - Further Detail} \label{nye_further}
The net Burgers vector $\bm{B}$ for a set of dislocations passing through a surface $\bm{S}$ can be defined by the surface integral
\begin{align} \label{eq:NyeTensorIntegral}
\bm{B} = \iint_{S} \bm{\alpha}^{Nye} \cdot d \bm{S}
\end{align}

One can apply Stokes' Theorem to the line integral from Eq.(\ref{eq:BurgersVector}), which results in the following surface integral \cite{rima};
\begin{align}
\bm{b} = \iint_{S} (\bm{\nabla} \times \bm{\beta})^{T} \cdot d \bm{S}
\end{align}
where $\bm{S}$ is the surface enclosed by Burgers circuit $\bm{C}$.

The resulting integrand is in fact the negative of the Nye tensor, $\bm{\alpha}^{Nye} = -(\bm{\nabla} \times \bm{\beta})^{T}$, which can be more easily interpreted as
\begin{align} \label{eq:NyeTensor}
\bm{\alpha}^{Nye} = \begin{bmatrix}
\pdv{\beta_{12}}{z}-\pdv{\beta_{13}}{y} & \pdv{\beta_{13}}{x}-\pdv{\beta_{11}}{z} & \pdv{\beta_{11}}{y}-\pdv{\beta_{12}}{x} \\
\pdv{\beta_{22}}{z}-\pdv{\beta_{23}}{y} & \pdv{\beta_{23}}{x}-\pdv{\beta_{21}}{z} & \pdv{\beta_{21}}{y}-\pdv{\beta_{22}}{x} \\
\pdv{\beta_{32}}{z}-\pdv{\beta_{33}}{y} & \pdv{\beta_{33}}{x}-\pdv{\beta_{31}}{z} & \pdv{\beta_{31}}{y}-\pdv{\beta_{32}}{x}
\end{bmatrix}
\end{align}
Note that the appearance of the minus sign is due to ambiguity caused by alternate conventions used to define the Burgers vector. The Nye tensor is actually related to the plastic displacement gradient (or slip tensor) $\bm{\beta}^{p}$ via $\bm{\alpha}^{Nye} = (\bm{\nabla} \times \bm{\beta}^{p})^{T}$ \cite{rima}, such that upon substitution into Eq.(\ref{eq:NyeTensorIntegral}) and applying Stokes' Theorem,
\begin{align}
\bm{B} = \oint_{C} \bm{\beta}^{p} \cdot d \bm{l}
\end{align}
However, it must be true that
\begin{align}
\oint_{C} (\bm{\beta} + \bm{\beta}^{p}) \cdot d \bm{l} = \bm{0}
\end{align}
as total displacement around a closed loop $\bm{C}$ must be zero \cite{rima}. Therefore it must be the case that $\bm{B} = -\bm{b}$; in other words the two expressions for the Burgers vector are opposite in direction. Since the sign of the Burgers vector is determined by the sign of the sense $\bm{\xi}$, which is itself arbitrary \cite{anderson,bacon}, the impact of the conflicting definitions is minor but should be borne in mind for the sake of clarity and consistency. In this paper Eq.(\ref{eq:BurgersVector}) as defined by Anderson et al. \cite{anderson} is the authoritative definition.

We shall now compute the Nye tensor for an infinite straight (singular) mixed dislocation. Consider the result for $\bm{\beta}$ found in Appendix \ref{derivations}\ref{mixed}. To find $\bm{\alpha}^{Nye}$ we populate Eq.(\ref{eq:NyeTensor}) with the elements of $\bm{\beta}$. The only non-zero terms are
\begin{align}
\pdv{\beta_{11}}{y} = \pdv{\beta_{12}}{x} = -\frac{b \sin(\alpha)}{4\pi(1-v)}\frac{1}{(x^2 + y^2)^3}\left[(3 - 2v)x^4 - 6x^{2}y^{2} - (1-2v)y^4\right], \\
\pdv{\beta_{21}}{y} = \pdv{\beta_{22}}{x} = -\frac{b \sin(\alpha)}{4\pi(1-v)}\frac{2xy}{(x^2 + y^2)^3}\left[(1+2v)x^2 - (3-2v)y^2\right], \\
\pdv{\beta_{31}}{y} = \pdv{\beta_{32}}{x} = \frac{b \cos(\alpha)}{2\pi}\frac{(y^2-x^2)}{(x^2 + y^2)^2}
\end{align}
and these terms all cancel, resulting in $\bm{\alpha}^{Nye} = \bm{0}$. Note that an exception to this is at the origin, as $\bm{\beta}$ is in fact non-differentiable there, so $\bm{\alpha}^{Nye}$ is instead undefined.

In fact, the Nye tensor is always undefined at the origin and $\bm{0}$ everywhere else for individual (singular) dislocations. This may at first seem counter-intuitive, however the reason is that the Nye tensor is evaluated at an infinitesimal point (as we are taking the curl of a tensor field, $\bm{\beta}$), and all of the discontinuities associated with the (singular) dislocation are confined to an infinitesimally thin line (i.e. the dislocation core). Essentially, the Nye tensor corresponds to having an infinitesimally small Burgers circuit. While this works mathematically across regions with dislocation presence defined by a density per unit area, it cannot work when we consider individual dislocations such that dislocation density is effectively zero everywhere and infinite at the dislocation cores.

Even for experimentally measured dislocations, deviations from linear elasticity theory will be limited to a small region around the dislocation core \cite{bacon}, thus using Nye tensor to determine the Burgers vector of an individual dislocation from experimental data is not feasible.

\section{Implementation} \label{implementation}
The program is split into a Plotter and a Calculator:
\begin{enumerate}
    \item The Plotter iterates through a chosen set of data and computes an approximate Burgers vector at every voxel/pixel position using a Burgers circuit of standard shape and size. The set of Burgers vectors can then be plotted as quivers in a 3D or 2D space, and the geometry of the material specimen itself, known dislocation lines, etc. can be superposed onto the plot if implemented by the user.
    \item The Calculator allows for manual input of Burgers circuit limits to target a dislocation or group of dislocations and accurately compute the Burgers vector therein.
\end{enumerate}

The two halves of the program can be used effectively in conjunction with each other, with the Plotter able to find dislocations and give estimates of their Burgers vectors and the Calculator then being used to investigate the dislocations in detail. The Calculator may also be configured to iterate through several concentric Burgers circuits to remove outliers and find an average computed Burgers vector.

The program can be adjusted to suit a wide variety of data; all it requires are the elastic strain and lattice rotation fields (in the form of evenly-spaced data arrays). The implementation of the data will naturally have to be adjusted depending on how the data is originally stored.

As well as working with experimental data, the program features a mathematical model for an infinite straight mixed dislocation with sense and Burgers vector being able to take any direction as described in appendix \ref{derivations}\ref{rotating}.

Further information can be found within the supplementary repository containing the program.

%%%%%%%%%% Insert bibliography here %%%%%%%%%%%%%%

%\bibliographystyle{RS} %%%% .BST file

%\bibliography{bibliography} %%%%% .Bib file

\end{document}